\documentclass[useAMS,usenatbib]{mn2e}

\usepackage{natbib}
\usepackage[dvipdfm]{graphicx}
\usepackage{amsmath,amssymb}

\def\ha{{\rm\,H$\alpha$}}
\def\hb{{\rm\,H$\beta$}}
\def\oi{{\rm\,[O{\sc i}]}}
\def\oii{{\rm\,[O{\sc ii}]}}
\def\nii{{\rm\,[N{\sc ii}]}}
\def\sii{{\rm\,[S{\sc ii}]}}

\def\oiii{{\rm\,[O{\sc iii}]}}
\def\hi{{\rm\,H{\sc i}}}
\def\hii{{\rm\,H{\sc ii}}}

\def\neiii{{\rm\,[Ne{\sc iii}]}}
\def\pa{{\rm\,Pa$\alpha$}}
\def\msun{{\rm M}$_{\odot}$}
\def\mn{\ifmmode {\mu {\rm m} ~}\else {$\mu $m ~}\fi}

\title[An early phase of environmental dependence of galaxy properties]
{An early phase of environmental effects on galaxy properties unveiled by near-infrared spectroscopy of protocluster galaxies at $\mathbf{z>2}$}
\author[R. Shimakawa, T. Kodama, K. Tadaki, M. Hayashi, Y. Koyama, and I. Tanaka]{Rhythm Shimakawa$^{1,2}$\thanks{E-mail:
rhythm@naoj.org (RS); t.kodama@nao.ac.jp (TK)}, Tadayuki Kodama$^{2,3}$\footnotemark[1], Ken-ichi Tadaki$^{3,4}$, Masao Hayashi$^{3}$,
\newauthor % starts a new line in the % author environment
Yusei Koyama$^{3,5}$, and Ichi Tanaka$^{1}$ \\
$^{1}$Subaru Telescope, National Astronomical Observatory of Japan, 650 North A$\textquoteright$ohoku Place, Hilo, HI 96720, USA \\
$^{2}$Department of Astronomy, School of Science, Graduate University for Advanced Studies, Mitaka, Tokyo 181-8588, Japan \\
$^{3}$Optical and Infrared Astronomy Division, National Astronomical Observatory, Mitaka, Tokyo 181-8588, Japan \\
$^{4}$Max-Planck-Institut f\"{u}r Extraterrestrische Physik, Giessenbachstrasse, D-85748 Garching, Germany \\
$^{5}$Institute of Space Astronomical Science, Japan Aerospace Exploration Agency, Sagamihara, Kanagawa, 252-5210, Japan}
\begin{document}

\date{Re-submitted 2014 November 25. Version 2.}

\pagerange{\pageref{firstpage}--\pageref{lastpage}} \pubyear{2014}

\maketitle

\label{firstpage}

\begin{abstract}
This work presents the results from our near-infrared 
spectroscopy of narrow-band selected \ha\ emitters (HAEs) 
in two rich overdensities (PKS 1138--262 at $z$=2.2 and 
USS 1558--003 at $z=2.5$) with the Multi-Object Infrared 
Camera and Spectrograph (MOIRCS) on the Subaru telescope.
These protoclusters are promising candidates for the most 
massive class of galaxy clusters seen today (Paper I).
The confirmed HAEs in the protoclusters at $z>2$ show high 
excitation levels as characterized by much higher \oiii/\hb\ 
or \oiii/\ha\ line ratios than those of general galaxies at 
low-$z$. 
Such a high excitation level may not only be driven by high 
specific star formation rates (sSFRs) and lower gaseous 
metallicities, but also be contributed by some other effects.
We investigate the environmental dependence of gaseous 
metallicities by comparing the HAEs in the protoclustrers 
with those in the general field at similar redshifts.
We find that the gaseous metallicities of protocluster 
galaxies are more chemically enriched than those of field 
galaxies at a given stellar mass in the range of 
M$_\star\lesssim10^{11}$ \msun. 
This can be attributed to many processes, such as intrinsic 
(or nature) effects, external (or nurture) effects, and/or 
some systematic sampling effects. 
The intrinsic (nature) effect leads to the advanced stage 
of `down-sizing' galaxy evolution in protoclusters.
On the other hand, the external (nurture) effects include
the recycling of chemically enriched gas due to the higher 
pressure of intergalactic medium and/or stripping of outer gas 
in the reservoir in protoclusters.
We also find that the offset of the mass--metallicity relation
in dense environment becomes larger at higher redshifts. 
This can be naturally understood by the fact that the 
inflow/outflow rates in star-forming galaxies are much higher 
at higher redshifts.
Therefore, the environmental dependence of such `feeding' and
`feedback' mechanisms in galaxy formation is probably playing 
major roles in producing the offset of the mass--metallicity 
relation for the protocluster galaxies at $z>2$.
\end{abstract}


\section{Introduction}
Local galaxy clusters ($z<1$) are dominated by red 
spheroidal galaxies and exhibit well-known relations, 
such as the colour--magnitude relation (red sequence) 
and morphology--density relation \citep{Butcher:1984, 
Dressler:1994, Dressler:1997, Kodama:2001}. 
The environmental dependence of such galaxy properties 
in the local Universe seems to originate from two effects, 
`nature' and `nurture'. 
The nature effect means that galaxies formed in dense 
environments are more evolved intrinsically due to 
biased galaxy formation, and thus are older hence redder 
\citep{Bower:1998, Thomas:2005}. 
In contrast, the nurture effect means that today's early-type 
galaxies in dense regions have been deformed through 
external effects from the surrounding environments 
\citep{Okamoto:2003, Prieto:2013}. 
For instance, galaxy mergers are expected to occur more 
frequently in high-redshift galaxy (proto-)clusters \citep{Gottlober:2001}.
In fact, dispersion dominated early-type galaxies are common 
in clusters (as reported by large kinematic studies such as 
ATLAS3D; \citealt{Cappellari:2011}). 
These may have formed via a major merger, which can 
produce a compact spheroidal galaxy through a dissipational 
kinematical process as predicted theoretically by many authors 
(e.g. \citealt{Mihos:1996, Bournaud:2007, Hopkins:2008}). 

In order to identify the physical processes that are 
responsible for the establishment of the environmental 
dependence, we need to trace back the history of galaxy 
formation and evolution in clusters. 
``Protoclusters" at $z\gg1$ are the ideal laboratories to 
investigate the progenitors of massive early-type galaxies in 
present-day rich clusters. 
\citet{Kodama:2007} showed that the tight red sequence at the 
bright end of luminosities breaks down in protoclusters in the 
redshift interval of $2<z<3$, which suggests that this epoch is 
a transition period of cluster galaxy formation. 
It is therefore essential to systematically explore properties 
of star-forming (SF) galaxies in protoclusters at $2<z<3$ in 
order to understand how the environmental dependence is 
established. 

With this motivation, several works have been devoted to 
the investigations of the environmental dependence of SF 
in protoclusters.
In particular, \citet{Kodama:2013} conducted a systematic 
study of protoclusters at $z>1.5$ with Subaru, as part of 
the `Mahalo-Subaru' ({\it Mapping HAlpha and Lines of Oxygen 
with Subaru}).
The Mahalo project targets 8 known clusters/protoclusters at 
$1.4<z<2.6$, and an un-biased field SXDS-UDS-CANDELS for 
comparison ($z=2.19$ and 2.53 slices).
We employ narrow-band (NB) filters on the two wide-field 
cameras, Subaru Prime Focus Camera (Suprime-Cam) and the 
Multi-Object Infrared Camera and Spectrograph (MOIRCS) on 
the Subaru telescope. 
Many NB filters are manufactured for the purpose of 
identifying \ha\ or \oii\ emitter candidates at that are 
physically associated with the clusters/protoclusters, as 
well as those located in narrow redshift slices in the field 
for comparison. 
The advantage of using the NB technique is that we can make 
a SFR limited sample with a high completeness, and the 
selection bias is minimized due to homogeneous, well-defined 
sampling using the star formation indicators.

It has been shown that the SF activity is high even in the 
cores of protoclusters at $z>2$, which scales as ($1+z$)$^{6}$ 
\citep{Shimakawa:2014}, or more rapidly \citep{Smail:2014, 
Shapley:2005b}. 
The peak of SF activity traced by line emitters is shifted 
from dense cluster cores to lower density outskirts and 
filamentary outer structures with time from $z\sim2.5$ to 
$z\sim0.4$, indicating the inside-out growth of clusters 
(e.g.\ \citealt{Hayashi:2012, Koyama:2013a, Hayashi:2011, 
Koyama:2010, Kodama:2004} in the order of decreasing redshift 
from 2.5 to 0.4). 

During such a turbulent, active epoch for galaxy formation, 
the physical conditions of inter-stellar medium (ISM) are 
expected to change. 
However, physical properties of line emitters in the 
protoclusters remain unclear so far. 
In the case of SF galaxies in general fields, for example, 
they form a tight sequence on the Baldwin, Phillips and 
Terlevich (BPT) diagram \citep{Baldwin:1981, Veilleux:1987} 
known as the {\it (chemical-)abundance sequence} 
\citep{Dopita:2000, 
Kauffmann:2003} and we can see more extremely ISM conditions 
at high-$z$ on this diagram as noted by many authors 
\citep{Shapley:2005, Erb:2006, Yabe:2012, Newman:2014, 
Kewley:2013a, Kewley:2013b, Nakajima:2013, Yabe:2014, 
Masters:2014, Steidel:2014, Shapley:2014, Coil:2014}. 
The abundance sequence of SF galaxies is shifted upward 
and/or rightward at high redshift (towards higher \oiii/\hb\ 
or \nii/\ha\ line ratios), and such a tendency is also seen 
in the field population at $z>2$ \citep{Erb:2006, Newman:2014, 
Kewley:2013a, Masters:2014}. 

In local galaxies, the ISM conditions are often described by 
some physical quantities such as ionization parameter ($q$), 
gaseous metallicity ($Z$) and electron density ($n_e$). 
At high-redshift, the ionization parameter is raised by a 
large flux of ionizing photons in ISM originated from hot O, B 
stars due to intensive star formation in relatively small 
galaxies. 
Previous studies suggest a high ionization parameter of SF 
galaxies at $z>2$ compared to that of local galaxies 
\citep{Erb:2010, Nakajima:2013, Nakajima:2014, Masters:2014}. 
Secondly, the chemical abundances of SF galaxies at $z\sim2$ 
is lower by 0.1--0.3 dex for a given stellar-mass compared to 
those at low-$z$ \citep{Erb:2006, Steidel:2014, Sanders:2014}. 
This leads to more compact and hotter O, B stars due to lower 
opacity \citep{Ezer:1971, Maeder:1987}, and thus UV radiation 
becomes harder and produces more ionizing photons. 
Thirdly, the strength of collisionally-excited emission lines 
(e.g. \oiii, \nii) strongly depends on the electron density. 
It is closely related to the number of electrons to collide 
since the excitation potential of this line is $\sim$1eV, 
which is nearly the same as the energy of electrons 
at the virial temperature ($\sim$10$^4$ K) \citep{Dyson:1980}.
Due to low excitation potential, the transition of 
collisionally-excited line is reliant on electron density 
compared to gaseous metallicity. 
Recent observations have suggested a high electron density 
($n_e>100$ cm$^{-3}$) in SF galaxies at $z\sim2$ 
\citep{Newman:2012, Masters:2014, Shirazi:2014, Wuyts:2014}. 
This value is larger than that of normal SF galaxies at low-$z$ 
by an order of magnitude, and close to that of interacting 
galaxies which are seen as (ultra) luminous infrared galaxies 
(ULIRGs/LIRGs) in the present-day Universe \citep{Krabbe:2014}. 
Such a large electron density contributes to the offset of 
galaxy distributions on the BPT diagram together with other 
physical parameters \citep{Brinchmann:2008b}. 
In this way, the cosmic dependence of the BPT diagram can 
be attributed to such physical parameters which determine 
ISM conditions. 

An obvious extension of such studies is to investigate the 
environmental dependence of the ISM properties at $z>2$ 
to test whether the evolutionary stages of SF galaxies, such 
as their chemical abundance, are environmentally dependent.
Protocluster galaxies may go through active nuclear starbursts 
followed by a quiescent phase due to higher chance of 
galaxy-galaxy mergers \citep{Gottlober:2001}.
Such nuclear starbursts would cause higher electron 
densities and harder UV radiation of the ISM  \citep{Shimakawa:2014c}, which 
lead to different gas excitation in the dense environments. 

Furthermore, we also want to address the environmental 
dependence of some characteristic relations of SF galaxies 
such as the mass--metallicity (M--Z) relation 
\citep{Tremonti:2004, Kewley:2008, Maiolino:2008, 
Yabe:2012, Zahid:2013}, since this relation reflects the star 
formation histories, as well as inflow/outflow processes in 
galaxies \citep{Erb:2006, Erb:2008, Zahid:2014, Yabe:2014b}. 
For example, if protocluster galaxies formed at the earliest 
times \citep{Thomas:2005}, the M--Z relation 
in the dense environments might be explained to be more 
evolved (i.e., offset to higher metallicity) compared to that 
in lower density regions at the same redshift. 

In this paper, we aim to explore the gaseous physical 
properties and its environmental dependence using 
near-infrared (NIR) spectroscopy of SF galaxies in 
two well studied protoclusters from the MAHALO-Subaru 
sample, namely PKS 1138--262 at $z=2.16$ 
\citep{Koyama:2013a} and USS 1558--003 at $z=2.53$ 
\citep{Hayashi:2012}, which are among the richest systems 
so far identified at $z>2$ \citep{Shimakawa:2014}. 
Since they both show large excesses in the number density 
of SF galaxies, these protoclusters are still in the vigorous 
formation process of present galaxy clusters.

At the redshift interval of $1.5<z<2.5$,
many important strong emission lines, including \ha, 
which are well-calibrated in local galaxies in the 
rest-frame optical, are all redshifted to the NIR regime 
($\lambda$=1.0--2.3 $\mu$m).
Our targets are selected from the NB \citep{Koyama:2013a, 
Hayashi:2012}, and thus NIR spectroscopy is efficient 
since the \ha\ emission is detectable at high signal-to-noise 
ratio (S/N) in a few hours.  

From the NIR spectroscopy, we investigate the following 
quantities of our \ha\ selected SF galaxies at $z>2$. 
We will measure the strength of dust attenuation using 
the Balmer decrement technique (\ha/\hb\ line ratios) 
and such as. 
We will study gaseous metallicities using the 
\citet{Pettini:2004} prescription (N2 index), excitation 
states, and the contribution (or contamination) of 
active galactic nucleus (AGN) using \nii/\ha\ versus 
\oiii/\hb\ diagram (BPT) \citep{Veilleux:1987} and stellar-mass
versus \oiii/\hb\ diagram (MEx) \citep{Juneau:2011}.
We then investigate the environmental dependence 
of the M--Z relation.

We first describe our data and the analyses.
We then show our results from various aspects; dust 
extinction (\S3.1), MEx diagram (\S3.2), BPT diagram 
(\S3.3), and M--Z relation (\S3.4). 
We discuss the differences of the physical properties 
and the SF activities between low versus high redshifts, 
and field versus dense environments. 
The final section gives the conclusions of this work. 

This work assumes the cosmological parameters of 
$\Omega_M$=0.3, $\Omega_\Lambda$=0.7 and 
$H_0$=70 km/s/Mpc. 
We employ the \citet{Salpeter:1955} initial mass function 
(IMF). 
When we show the results in the literature that use the
\citet{Chabrier:2003} IMF, we scale up SFRs and stellar masses
by a factor of 1.8 to match them to the case of the Salpeter IMF.

%%%%%%%%%%%%
%% Methodology %%
%%%%%%%%%%%%
\section{Data \& analysis}

\subsection{Observational data}

%%%%%%%%%%%%%
\begin{figure}
\centering
\includegraphics[width=85mm]{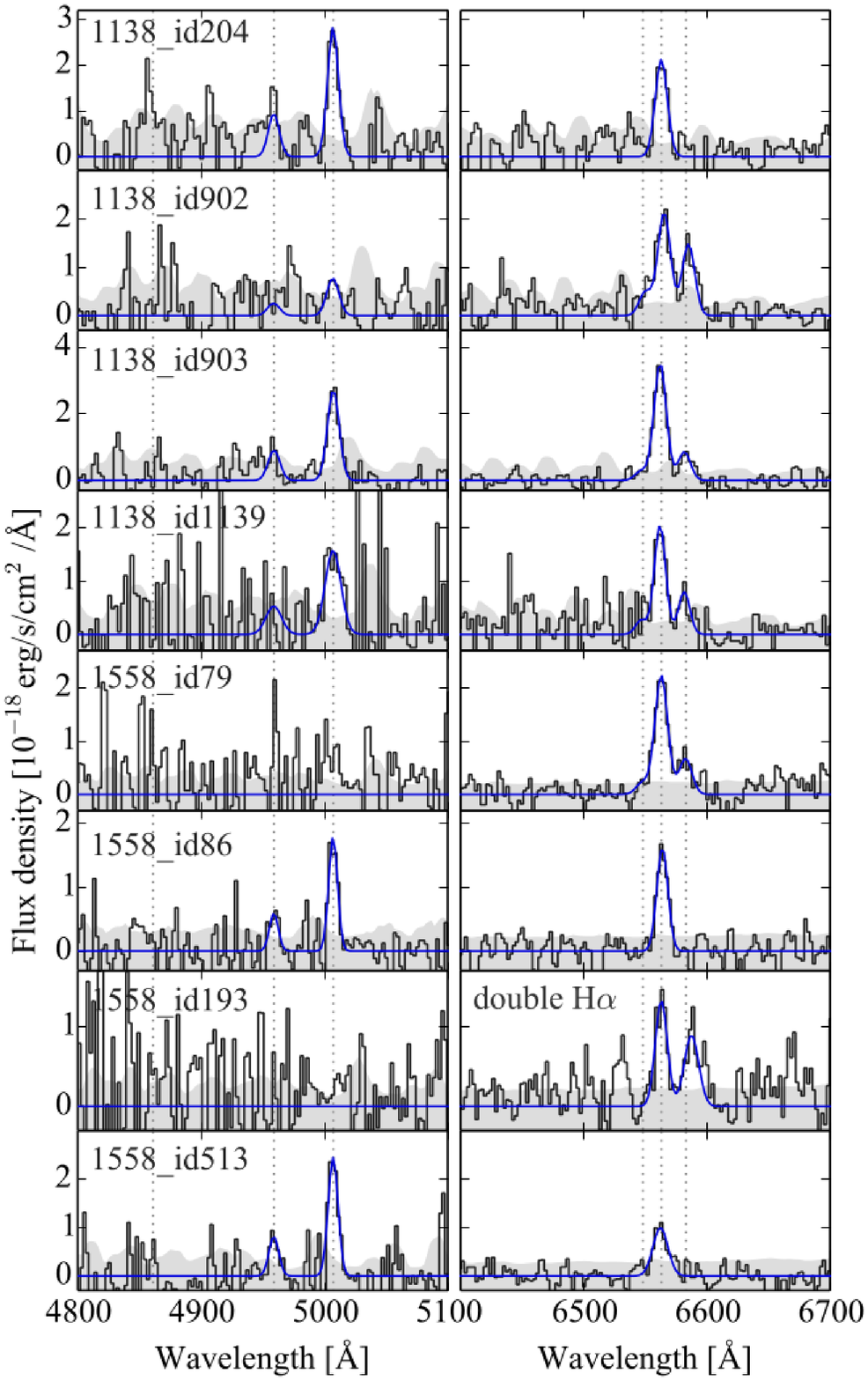}
\caption{Examples of the spectra of HAEs in the 
PKS1138 and USS1558 protoclusters.
The black, blue line and grey filled region show reduced spectra, 
fitting curve, and sky Poisson noise levels. 
We here subtract a continuum of each spectrum. 
The dotted vertical lines show \nii6583\AA\, \ha6563\AA\, 
\nii6548\AA\, \oiii5007\AA\, \oiii4959\AA\, \hb4861\AA\ 
emission lines from the right side.}
\label{fig1}
\end{figure}
%%%%%%%%%%%%%

In this paper, we study the physical properties of HAEs in 
the two protoclusters associated with radio galaxies, 
PKS 1138--262 ($z=2.16$; hereafter PKS1138) and 
USS 1558--003 ($z=2.53$; hereafter USS1558).
In the respective protoclusters, 48 and 68 HAE candidates 
that show excesses of NB fluxes above 3$\sigma$ threshold 
and are also identified as cluster memberships from a 
broad-band colour-colour diagram ($Bz^{\prime}K_s$ or $r^{\prime}JK_s$) 
to separate \ha\ emitters (HAEs) at the cluster redshift 
from contaminant \oiii/\oii/\pa\ emitters at other redshifts 
(see \citealt{Koyama:2013a, Hayashi:2012} for more details). 

The spectroscopic observations of these candidates were
conducted in 2013 April with MOIRCS, a NIR imager and 
spectrograph \citep{Ichikawa:2006, Suzuki:2008} mounted 
on the 8.2-m $Subaru$ telescope on Mauna Kea. 
We employed 3 slit masks for each protocluster, and a 
typical slit mask includes about 20 science targets. 
The net integration time was 2 to 3 h per mask, and the 
typical seeing size was about 0.7 arcsec. 
In total 98 HAE candidates were observed by using the 
low resolution grism (HK500: $\lambda$=1.3--2.5 $\mu$m, 
$R$$\sim$500 for 0.8 arcsec slit width) for 5 masks, and 
the high-resolution grism (VPH-K: $\lambda$=1.9--2.3 $\mu$m, 
$R$$\sim$1700 for 0.8 arcsec slit width; see 
\citealt{Ebizuka:2011}) for one of the three masks of PKS1138. 
A summary is given in Table 1 (see also \citealt{Shimakawa:2014}).

%%%%%%%%%%%%%
\begin{table*}
\centering
\caption{
A summary of data-sets obtained and used in this work.
Columns: (1) protocluster name, (2) redshift, (3) the number 
of HAE candidates selected by the combined technique of NB 
and broad-band \citep{Koyama:2013a, Hayashi:2012}, (4) 
3$\sigma$ limit of SFRs (without dust correction) of HAEs 
in the past studies, (5) grism name, (6) spectral resolution 
with 0.8 arcsec slit width, (7) integration time, (8) the 
number of science objects in each slit mask, (9) the number 
of total observed targets by low resolution grism, HK500 
(including VPH-K), and (10) the number of confirmed cluster 
members by HK500 (including VPH-K).}
\begin{tabular}{@{}cccccccccc@{}}
\hline
Protocluster & $z$ & HAE & SFR$_\mathrm{lim}$ & Grism & R & Int. time & Slit & Target & Confirmation \\
 (1) & (2) & (3) & (4) & (5) & (6) & (7) & (8) & (9) & (10) \\
\hline
PKS 1138--262 & 2.156 & 48 & $\sim$10 \msun/yr & HK500 & 513 & 120 min & 23 & 36 (44) & 24 (27) \\
($\alpha_\mathrm{J2000}$, $\delta_\mathrm{J2000}$) & & & & HK500 & 513 & 161 min& 19 & & \\
(175.2013, --26.4858) & & & & VPH-K & 1675 & 225 min & 18 & & \\
\hline
USS 1558--003 & 2.533 & 68 & $\sim$8 \msun/yr & HK500 & 513 & 180 min & 25 & 54 & 36 \\
($\alpha_\mathrm{J2000}$, $\delta_\mathrm{J2000}$) & & & & HK500 & 513 & 276 min & 19 & & \\
(240.3221, --0.4797) & & & & HK500 & 513 & 175 min & 15 & & \\
\hline
\end{tabular}
\label{tab1}
\end{table*}
%%%%%%%%%%%%%

We reduced the obtained spectra using the MOIRCS 
MOS Data Pipeline ({\sc mcsmdp}: \citealt{Yoshikawa:2010}) 
that is an {\sc IRAF} based data reduction software for 
spectroscopic data of MOIRCS \footnotemark[1].
This reduction package was designed to operate standard 
procedures semi-automatically which consists of bad pixel 
masking, cosmic rays filtering, A$-$B subtraction, 
wavelength calibration, distortion correction, and 
background subtraction (more detail is described in 
\citealt{Yoshikawa:2010}). 
Then, we conducted a flux calibration based on a standard 
star with a A0V spectrum and the K-band magnitude of 9.930. 
Reduced spectra show the mean 1$\sigma$ sky 
noise of $m_{AB}$=22.2--22.6 and the root-mean-square 
error (rms) of wavelength calibration of about 
$\Delta v$=$\pm$40--50 km/s. 
Firstly, we looked for \ha\ emission lines at around 
the protocluster redshifts within $\pm$2000 km/s, and 
performed a Gaussian fitting using the software 
{\sc specfit} \citep{Kriss:1994} distributed within 
{\sc stsdas}\footnotemark[2] layered on top of the 
{\sc iraf} environment. 
We usually used a single Gaussian fitting iteratively, 
but sometimes applied a multi-Gaussian fitting for a broad 
or a multiple blended emission line, and then the chi-square 
minimization technique was used to best fit the line profile. 
We here included sky Poisson noises, $\sigma(\lambda)$, as 
a parameter in the line fitting. 
The 1$\sigma$ flux errors discussed throughout this paper 
were produced by this process, and hence they include both 
fitting errors and sky noise.
\footnotetext[1]{{\sc iraf} is distributed by National Optical Astronomy 
Observatory and available at iraf.noao.edu/}
\footnotetext[2]{Available at www.stsci.edu/institute/software\_hardware/stsdas/}

We identify \ha\ emission lines at above 3$\sigma$ 
levels for 28 and 37 emitters in PKS 1138 and USS 1558 
protoclusters, respectively. 
However, each one of them have been identified as 
background \oiii\ emitter since it has doublet lines 
\oiii$\lambda$4959/5007 in each protocluster region. 
Also, we could not find any objects which have emission 
lines only outside the wavelength of the response curves 
of the NB filters. 
The remaining 37 non-detections tend to have lower NB 
fluxes or larger stellar masses, which indicates that our 
spectroscopy is not sensitive enough to detect faint or 
broad emission lines.

This work adopts confirmed HAEs detected by the 
low resolution grism (Table \ref{tab1}), 
since we can discuss the physical properties inclusive 
of \oiii\ and \hb\ lines. 
Therefore, final sample is composed of 24 and 36 HAEs 
in PKS 1138 and USS 1558, respectively. 
While about a half of the resultant spectra show \oiii\ 
lines coupled with \ha\ lines with above 3 sigma level, 
\nii\ and \hb\ lines cannot be detected in most of the sample. 
Some examples of the resultant spectra including best fit 
curves, and sky Poisson noise levels are shown in 
Fig. \ref{fig1}. 

In this work, we often use the NB flux, stellar-mass, 
and optical band magnitude derived from the previous 
imaging surveys with MOIRCS and Suprime-cam 
\citep{Koyama:2013a, Hayashi:2012}. 
For comparison of our data with field counterparts 
at the similar redshifts, we are referring several samples 
as reported by \citet{Newman:2014, Erb:2006} 
These data sets are based on UV colour \citep{Adelberger:2004}. 
This may cause significant descrepancies from our 
\ha-selected sample. 
We discuss the effects of sampling bias in the section 4.2. 
Also, we use SDSS sample \citep{Abazajian:2009} at 
$0.04<z<0.3$ as comparison of the galaxies at between 
low redshift and high redshift, and we cite their stellar 
masses and SFRs from \citet{Kauffmann:2003, Salim:2007} 
and \citet{Brinchmann:2004}, respectively. 
All of those IMF are corrected to the \citet{Salpeter:1955} 
IMF. 
Through this paper, we select the SDSS galaxies with 
equivalent width (EW) of \ha\ emission line greater than 
20\AA, because our data are based on the previous NB 
imaging \citep{Koyama:2013a, Hayashi:2012} which 
selects the HAEs with EW$_\mathrm{H\alpha}>$20\AA\ 
in the rest frame. 
Moreover, we select only those whose \nii$\lambda$6583\AA, 
\ha$\lambda$6563\AA, \oiii$\lambda$5007\AA\, and 
\hb$\lambda$4861\AA\ emission lines are all detected more 
than 3$\sigma$ reliability.

\subsection{Stacking analysis}

27 and 36 HAEs in PKS1138 and USS1558 have been confirmed. 
For 13 and 17 confirmed HAEs in respective protoclusters, we 
also detect \oiii\ lines. 
As shown in the next section, we study their physical properties 
on the \oiii/\hb\ line ratio versus stellar-mass plane based on 
their \oiii/\ha\ line ratios by assuming an appropriate dust 
correction from their UV/\ha\ luminosity ratios. 
It is not ideal to discuss the physical properties of the 
protocluster galaxies with only \ha\ and \oiii\ lines.
In most of the individual spectra, however, do not detect 
\nii\ or \hb\ emission lines above a S/N=3. 
In order to study physical properties of protocluster galaxies 
such as metallicity and excitation and compare them with 
those of field galaxies at the same redshifts, 
we need four emission lines (\ha, \oiii, \nii, \hb). 

We thus apply a stacking analysis by summing 
the individual spectra.
This technique is useful to investigate `average' properties. 
Here, we stack the individual spectra using the S/N as the weight. 
We divide our sample into two bins by stellar mass in each 
protocluster with approximately equal numbers per bin
as shown in Table 2 (PKS1138-low, -high and USS1558-low, -high). 
In the stacking analysis, we only use the spectra taken by the 
low-resolution grism HK500. 
The spectral stacking is done using the following equation, 
\begin{equation}
F_{stack}(\lambda) = \sum_i^n\frac{F_i(\lambda)}{\sigma_i(\lambda)^2} / \sum_i^n\frac{1}{\sigma_i(\lambda)^2},
\end{equation}
where $F_i(\lambda)$ is a flux density of an individual 
spectrum and $\sigma_i(\lambda)$ is a sky poisson noise 
as a function of wavelength. 
We then apply the least chi-square fitting technique to the 
stacked line profiles by multi Gaussian curves with {\sc specfit} 
as we apply to the individual spectra. 
We sum up the spectra with a `median' combine, as it 
produces a spectrum with the highest S/N as it suppresses 
the sky line residuals. 

%%%%%%%%%%%%%
\begin{table}
\begin{center}
 \caption{Properties of the stacked spectra of \ha\ emitters 
 in PKS1138 and USS1558 separated into two stellar 
 mass bins (low and high) and the same stellar-mass bin for 
 comparing with each protocluster (mid). 
 (1) ID categorized by stacking method, (2) number of galaxies 
 in each bin, (3) the median stellar-mass, 
 1$\sigma$ scatter and (4) stellar-mass range of individual 
 spectra used in stacking. }
 \begin{tabular}{@{}cccc@{}}
\hline
  ID & Num. & log(M$_\star$/M$_\odot$) & log(M$_\star$/M$_\odot$) range \\
  (1) & (2) & (3) & (4) \\
\hline
PKS1138-low & 13 & $9.99_{-0.73}^{+0.43}$ & 8.60--10.54 \\
PKS1138-high & 9 & $11.16_{-0.34}^{+0.20}$ & 10.68--11.56 \\
\hline
USS1558-low & 15 & $9.66_{-0.48}^{+0.32}$ & 8.98--10.08 \\
USS1558-high & 11 & $10.64_{-0.24}^{+0.19}$ & 10.21--11.07 \\
\hline
PKS1138-mid & 11 & $10.15_{-0.19}^{+0.22}$ & 9.86--10.54 \\
USS1558-mid & 7 & $10.25_{-0.23}^{+0.20}$ & 9.83--10.56 \\
\hline
\end{tabular}
\end{center}
\label{tab2}
\end{table}
%%%%%%%%%%%%%
%%%%%%%%%%%%%
\begin{figure}
\centering
\includegraphics[width=80mm]{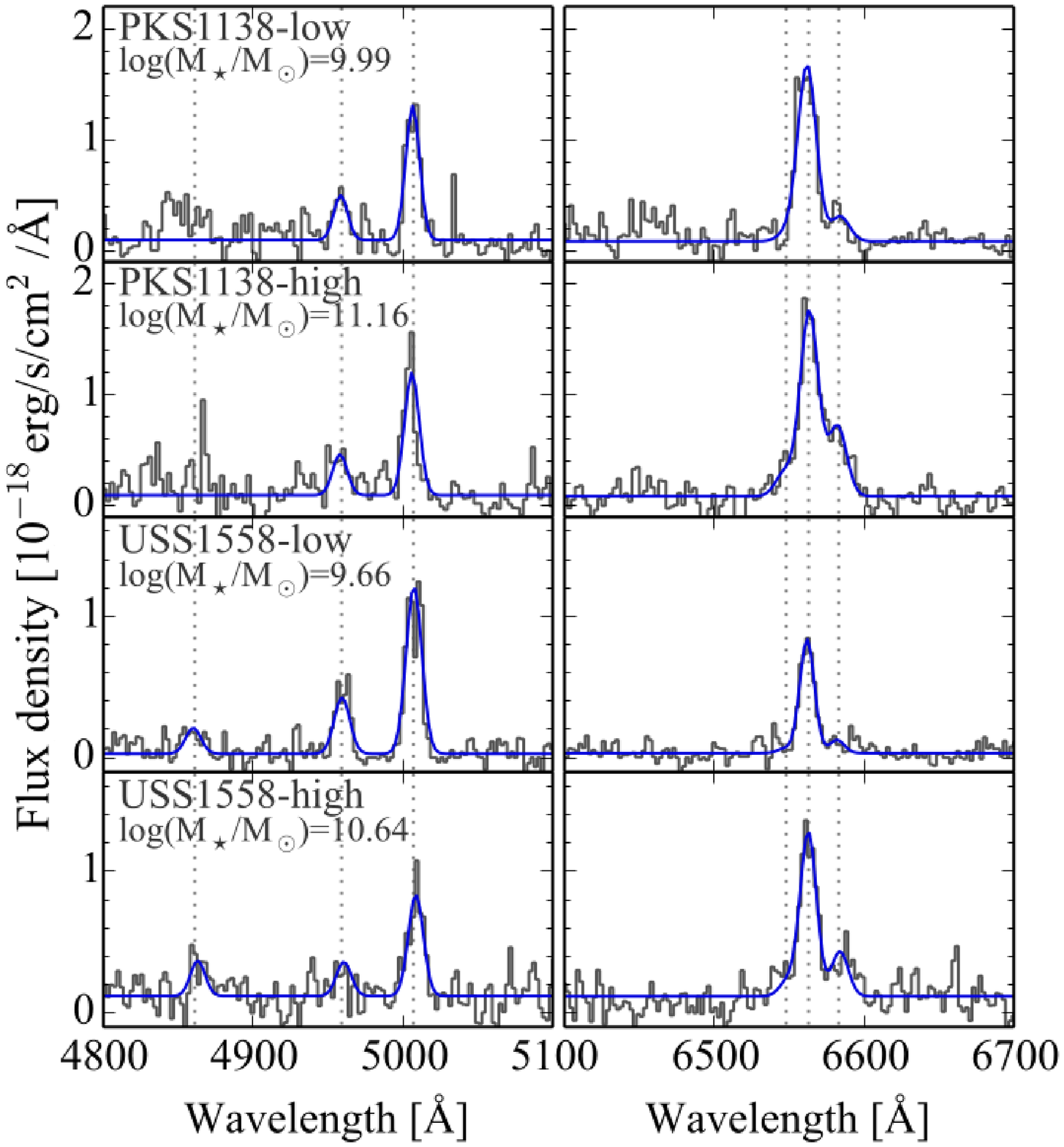}
% \vspace*{10cm}
\caption{ 
The resultant stacked spectra (from top to bottom: 
PKS1139-low, -high, USS1558-low, and -high). 
Grey and blue curves show stacked spectra and the 
results of the spectral fit.
Dotted vertical lines are the same as those in
Fig. \ref{fig1}.}
\label{fig2}
\end{figure}
%%%%%%%%%%%%%

The spectral fitting results are shown in Fig. \ref{fig2}. 
We see detections of \nii\ and \hb\ lines (although at low 
S/N) in most of the stacked spectra. 
Also, clear mass dependences of \nii, \oiii, and Balmer lines 
can be found in the sense that the stacked spectra of high 
stellar mass bins tend to have stronger \nii\ 
and weaker \oiii\ compared to those of lower mass bins 
(see also \citealt{Shapley:2014}). 
In addition, for a direct comparison between the two 
different protoclusters, we also stack the spectra with a 
common stellar mass bin, and they are represented
by PKS1138-mid and USS1558-mid. 
However, we cannot find any difference between the two 
observed protoclusters in the stacked spectra of the same 
stellar-mass bin (Table 2). 
Therefore we hereafter only show the stacked spectra of 
PKS1138-low, -high, USS1558-low, and -high, to improve 
visibility of the figures, unless otherwise noted.

%%%%%%%%%%%%
%%     Results     %%
%%%%%%%%%%%%
\section{Physical States of SF Galaxies}

The physical properties of SF galaxies in dense environments 
may provide us with information on how those galaxies are 
formed, and affected by their surrounding environments (such 
as galaxy-galaxy mergers), and how they evolve together with 
central supermassive black holes. 

\subsection{Dust extinction}

We here investigate the dust attenuation of the protocluster 
galaxies at $z>2$ in two ways, 
first, based either on the Balmer decrement or second on the 
ratios between SFR(\ha) and SFR(UV) \citep{Kashino:2013, 
Wuyts:2013, Koyama:2014} by assuming the \citet{Calzetti:2000} 
extinction curve, $k(\lambda)$. 

In the former case, we estimate the amount of dust extinction 
for \ha\ flux using the observed
\ha/\hb\ line ratio measured from the stacked spectra, 
\begin{equation}
A_{\mathrm{H\alpha,Balmer}} = 6.535\ \log(F_\mathrm{obs}(\mathrm{H\alpha})/F_\mathrm{obs}(\mathrm{H\beta})) -2.982
\label{eq2}
\end{equation}
We here assume that the intrinsic \ha/\hb\ line flux ratio 
is 2.86 for a Case B recombination in the gas temperature 
of T$_e$=10$^4$ K and the electron 
density of $n_e=10^2$ cm$^{-3}$ \citep{Brocklehurst:1971}. 

In the latter case, the extinction magnitude for a \ha\ line 
can be estimated by the following equation:
\begin{equation}
A_\mathrm{H\alpha,UV} = A_\mathrm{UV} + 2.5\ \log(\mathrm{SFR_\mathrm{obs}(UV)/SFR_\mathrm{obs}(H\alpha)}),
\label{eq3}
\end{equation}
where SFR(UV) is estimated by using the \citet{Kennicutt:1998} 
conversion and SFR(UV)=1.4$\times$10$^{-28}$ $L_\nu$ erg/s/Hz. 
We calculate $L_\nu$ from the $B$-\footnotemark[3] and $r'$- 
band fluxes ($\lambda_\mathrm{rest}$$\sim$1400\AA\ and 
1770\AA\ for PKS1138 and USS1558, respectively). 
According to the \citet{Calzetti:2000} prescription for dust 
extinction, $k(\lambda)$ values at three relevant wavelengths 
are $k(\lambda$6563\AA)=$k_\mathrm{H\alpha}$=3.33; 
$k(\lambda$1400\AA)=10.78; $k(\lambda$1770\AA)=9.48. 
The actual absolute values of extinction depend also on the 
effective obscuration factor, $R_V$=$A_V$/$E_s$$(B-V)$. 
This work adopts $R_V$=4.05 \citep{Calzetti:2000}. 
SFR$_\mathrm{obs}$(\ha) is also estimated by the 
\citet{Kennicutt:1998} calibration, 
SFR$_\mathrm{obs}$(\ha)=7.9$\times$10$^{-42}$$L_\mathrm{H\alpha}$, 
where $L_\mathrm{H\alpha}$ is the \ha\ line luminosity. 
We use the \ha\ flux estimated from the NB flux ($F$(NB)) 
to avoid any flux loss.
The NB fluxes are sampled by our previous works 
\citep{Koyama:2013a, Hayashi:2012}. 
It should be noted that $F$(NB) usually contains \nii\ 
emission line flux, and we need to correct for such 
contamination. 
Here we apply a mass-dependent flux correction. 
The correction is determined as a function of stellar-mass 
using the correlation between stellar-mass and \nii/\ha\ 
line ratio that we derive from the stacking analysis,
which is equivalent to the mass--metallicity relation 
(sees more detail in \S3.4). 
\footnotetext[3]{The B-band data of PKS1138 field was 
originally published by \citet{Kurk:2000, Kurk:2004a}). 
Source photometry was performed by \citet{Koyama:2013a} 
at the positions of our HAEs, and in this paper we use 
their catalogued magnitudes.}

Furthermore, we need to convert from stellar absorption 
to nebular attenuation 
($A_\mathrm{nebula}$$\propto$$A_\mathrm{star}$/$f$). 
The strength of dust obscuration depends not only on 
dust-to-gas ratio but also on the geometry of dust and OB 
stars in \hii\ regions where the emission lines originate.
We explore an appropriate extra extinction of nebular 
emission lines compared to that of stellar light since it is 
critical for studying physical properties of SF galaxies as 
discussed by many authors \citep{Yoshikawa:2010, 
Kashino:2013, Wuyts:2013, Whitaker:2014}. 
\citet{Kashino:2013} have reported that $f$ value can be 
larger for distant SF galaxies at $z>1$ ($f$=0.69--0.83) 
based on $sBzK$-selected galaxies \citep{Daddi:2007a}. 
On the other hand, if the cloud scaling relation reported by 
\citet{Larson:1981} breaks down at high redshifts, the 
proportionality would change \citep{Wuyts:2013}. 
\citet{Wuyts:2013} have proposed a new approach for dust 
correction, which uses a non-linear conversion from stellar 
to nebular extinctions based on 473 massive SF galaxies 
(M$_\star>10^{10}$ \msun) at $0.7<z<1.5$ who derive 
$A_\mathrm{nebula}$=$A_\mathrm{star}$(1.9$-$0.15$A_\mathrm{star}$). 

%%%%%%%%%%%%%
\begin{figure}
\centering
\includegraphics[width=80mm]{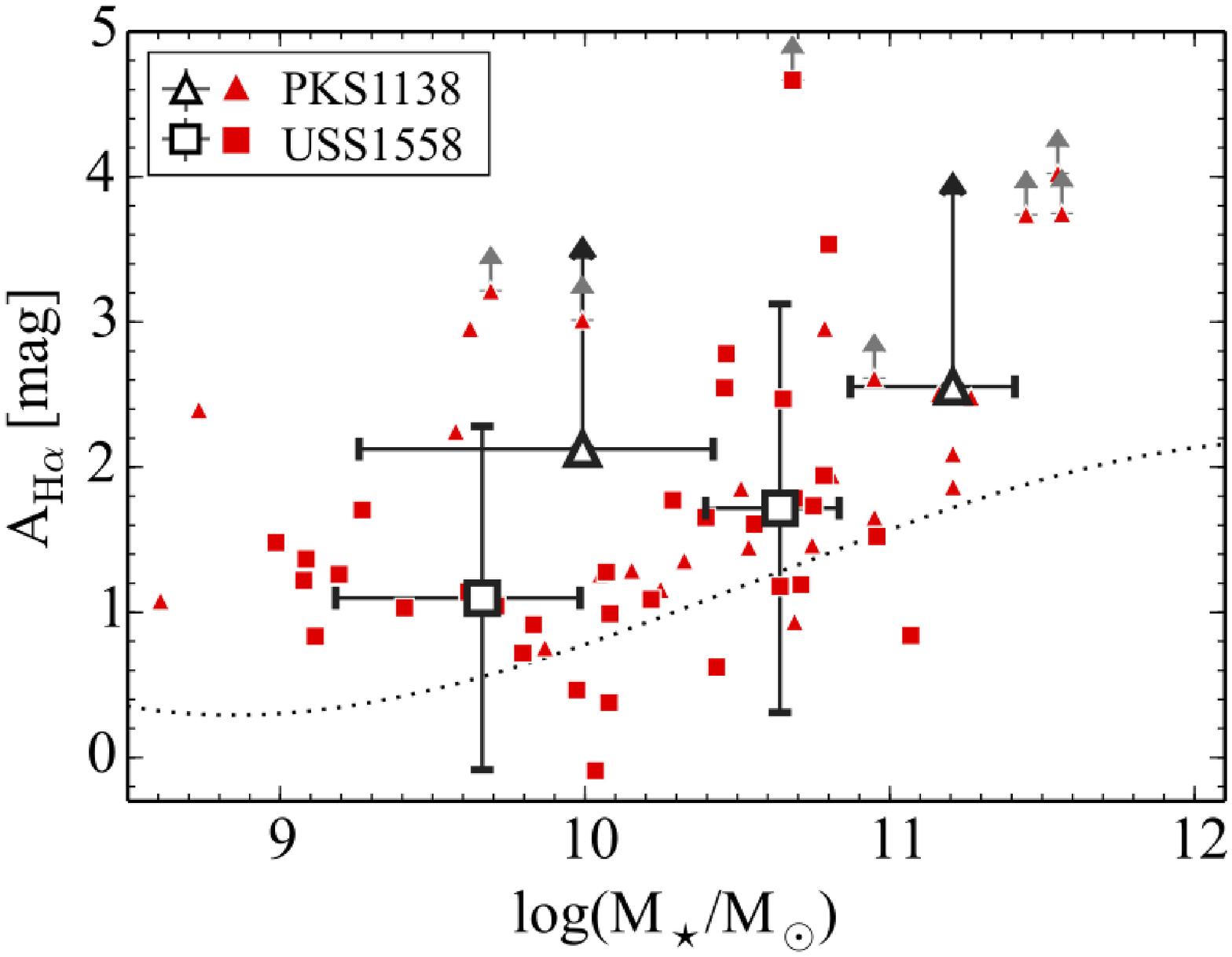}
\caption{Dust extinction of \ha\ lines as a function of 
stellar mass. 
Large open triangles and squares indicate the stacked spectra 
of HAEs in PKS1138 and USS1558, respectively, and their 
error-bars represent $\pm$ 1 $\sigma$ errors.
Red filled triangles and squares show the individual HAEs in 
PKS1138 and USS1558, respectively. 
For the objects whose detection significance is lower than 
2$\sigma$ in the B-band, we use the 2$\sigma$ limiting 
magnitude in B or r'-band to estimate the UV flux.
We also use the 2$\sigma$ flux if the \hb\ flux of a stacked 
spectrum is less than the 2$\sigma$ detection limit to estimate 
$A_\mathrm{H\alpha,Balmer}$.
Dotted curve is an empirical sequence of SDSS galaxies reported 
by \citep{Garn:2010}.}
\label{fig3}
\end{figure}
%%%%%%%%%%%%%

We investigate an appropriate $f$ value which can 
give consistent dust attenuation values for HAEs in USS1558
with both the Balmer decrement technique and the UV/\ha\ 
method. 
After trial and error attempts, a linear conversion factor is
found to be more favorable
for our sample since the \citet{Wuyts:2013} prescription 
shows much higher $A_\mathrm{H\alpha,UV}$ (0.5--1.0 dex) on 
average compared to $A_\mathrm{H\alpha,Balmer}$ of the 
stacked data. 
This paper uses the best value of $f$=0.66.
Figure \ref{fig3} shows $A_\mathrm{H\alpha}$ versus 
stellar-mass relation. 
With the $f$ value and the prescrption as described above,
the discrepancy, $\Delta A_\mathrm{H\alpha}=
A_\mathrm{H\alpha,Balmer}-A_\mathrm{H\alpha,UV}$
gets as small as $-0.021$ for HAEs with 
M$_\star$$<$10$^{10.1}$\msun\ (USS1558-low) and $+0.016$ 
for those with M$_\star$$>$10$^{10.2}$\msun\ (USS1558-high). 
From now on, we ignore such small residuals as they 
have no impact on our results.
In Fig. \ref{fig3}, local relationship for the SDSS galaxies 
is also shown by a dotted curve \citep{Garn:2010}. 
While we do not see any significant difference in \ha\ 
extinction of our sample compared to the local sequence, 
we identify 7 heavily-obscured objects with 
$A_\mathrm{H\alpha}>3$ in the protoclusters.
In particular, one of them seen in USS1558 that shows a 
significant excess from other galaxies ($A_\mathrm{H\alpha}>4$) 
has a total IR luminosity of $5\times10^{12}$L$_\odot$ and 
its radio continuum is detected by the Jansky Very Large Array 
\citep{Tadaki:2014b}. 
More details regarding the dusty and starburst populations 
and their environmental dependence will be discussed in a 
forthcoming paper.

\subsection{Mass--excitation diagram}

Various line ratios reflect physical properties, such as 
gaseous metallicity, ionization parameter, and electron 
density of the ionized gas in \hii\ regions. 
Line ratios are usually measured for sets of strong lines 
such as \oiii/\hb\ and \nii/\ha, located nearby in wavelength 
so that the ratio is almost free from dust extinction.
In order to derive the physical quantities from the line 
ratios, many authors have improved the line diagnostics 
using theoretical models and/or the empirical calibrations 
based on local galaxies. 
Among them, some traditional treatments are widely used, 
such as BPT diagram (\citealt{Baldwin:1981}, see \S3.3) and 
mass-metallicity relation (\citealt{Tremonti:2004}, see \S3.4).
In \S3.2--3.4, we show the physical properties of HAEs in 
the protoclusters based on several line diagnostics, and 
compare them with SDSS local galaxies and the field galaxies 
at similar redshifts ($z$=2.1--2.5). 

This MOIRCS spectroscopy not only confirms the existence 
of \ha\ emission line for a large number of NB-selected HAE 
candidates, but also detects \oiii\ emission line for about a 
half of the confirmed HAEs. 
Interestingly, their \oiii\ fluxes are often higher than their 
\ha\ fluxes. 
The fraction of \oiii\ detection (50\%) is much larger by 
about 3 times than expected from the typical \oiii/\ha\ line 
ratio of local SF galaxies. 
In the local SDSS galaxies that we select in this paper, 
for example, $\sim$97\% of the HAEs (i.e. EW$>$20\AA) 
have \oiii\ line fluxes that are smaller than \ha\ fluxes 
(without dust correction) even including Seyfert AGNs, 
while our HAE sample at $z$$>$2 has higher \oiii\ fluxes 
than \ha\ at a rate of $>$43\% with reddening correction. 
This may indicate that the SF galaxies at $z>2$ have 
much higher ionization states in dense regions 
the same as in the field environments (e.g. 
\citealt{Kewley:2013a, Shirazi:2014, Steidel:2014}). 

By studying the relationship between \oiii/\ha\ line ratio 
and stellar mass, we obtain a diagnostic which can separate 
Seyfert AGNs from SF galaxies. 
This diagram is called Mass-Excitation (MEx) diagram which 
was first discussed by \citet{Juneau:2011}. 
The original MEx is presented by using a close-pair line 
ratio, \oiii/\hb. 
However, this work alternatively employs the \oiii/\ha\ 
line ratio for individual objects to derive \oiii/\hb\ 
ratio by assuming the dust extinction (UV/\ha) as presented 
in the previous section, because \hb\ lines are rarely 
detected for individual galaxies.

%%%%%%%%%%%%%
\begin{figure}
\centering
\includegraphics[width=80mm]{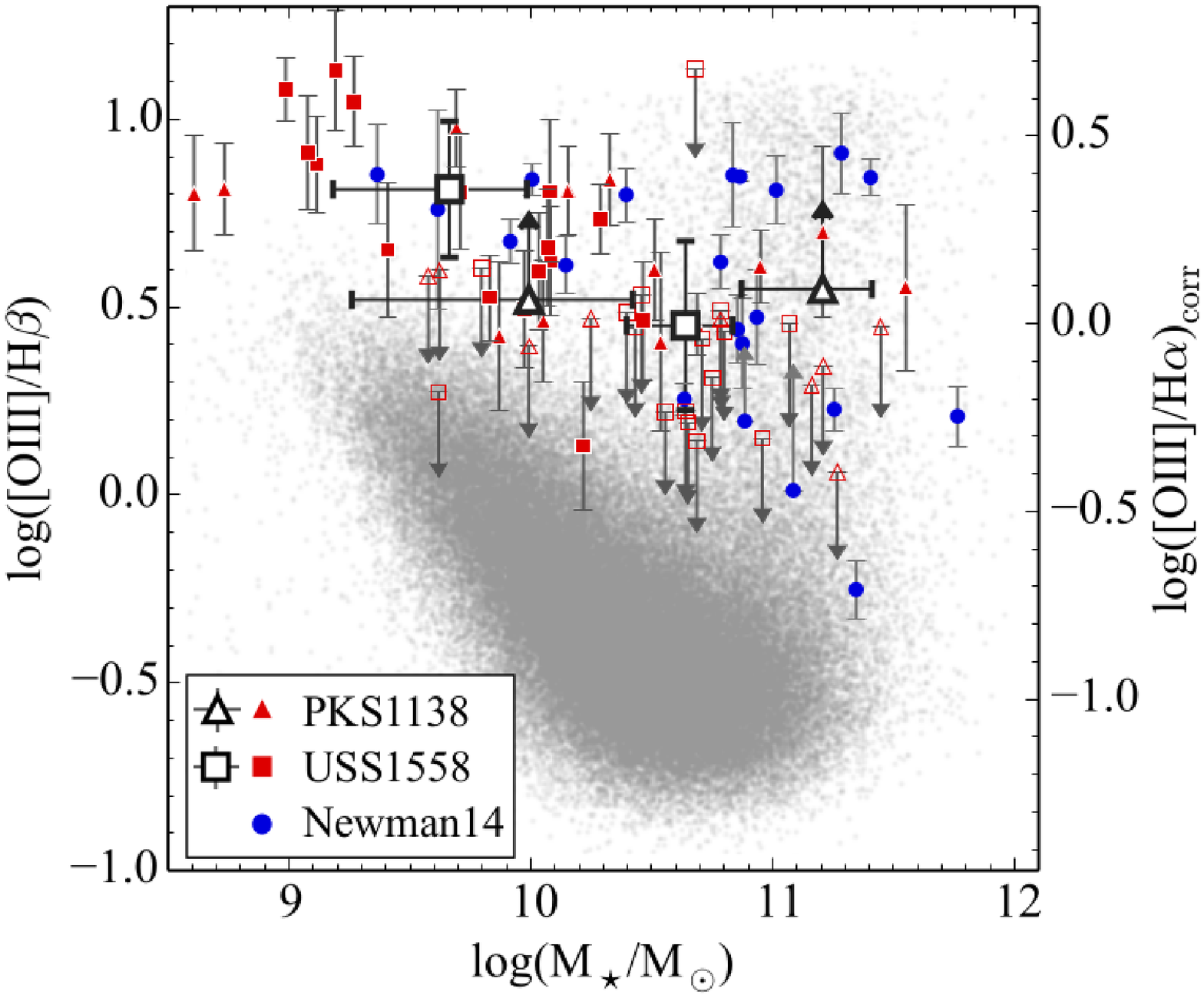}
\caption{Mass-excitation (MEx) diagrams for individual and 
stacked results. 
White triangles and squares indicate the stacked spectra of 
in PKS1138 and USS1558, respectively.
Red triangles, squares, and blue circles show the individual 
HAEs in PKS1138, USS1558, and SINS \&\ LUCI data (field) 
at $z$=2.1--2.5 \citep{Newman:2014}, respectively. 
\oiii/\hb\ line ratio of individual objects in PKS1138 and 
USS1558 is based on the prescriptions of \citet{Wuyts:2013} 
and intrinsic \ha/\hb\ flux ratio (=2.86). 
We show the 2$\sigma$ upper limit for \oiii-undetected 
($<2\sigma$) objects. 
\hb\ absorption is not considered for the \citet{Newman:2014} 
sample. 
Grey dots represent the SDSS galaxies with 
EW$_\mathrm{H\alpha}$$>$20\AA\ \citep{Abazajian:2009}. 
Error-bars indicate flux errors of 1$\sigma$ in the respective samples. 
}
\label{fig4}
\end{figure}
%%%%%%%%%%%%%

Figure \ref{fig4} shows the MEx diagram. 
We can see that the \oiii/\hb\ line ratios (from \oiii/\ha) 
of our sample are higher by an order of magnitude than 
SDSS galaxies at the same stellar mass. 
This indicates that the ionizing states of high-$z$ galaxies are 
significantly higher than low-$z$ galaxies. 
Such high excitations of the protocluster galaxies at $z>2$ are 
consistent with those of field galaxies as noted by recent work 
\citep{Newman:2014, Holden:2014, Juneau:2014, Coil:2014}. 
The success rate of \oiii\ detection ($>2\sigma$) is poor for 
high-mass galaxies and the 2$\sigma$ limit flux is assigned 
to the objects if \oiii\ line is not detected. 
The absence of the \oiii\ line in high-mass galaxies is likely 
to be caused by both influence of higher dust extinction 
\citep{Garn:2010} and intrinsically smaller \oiii\ fluxes 
\citep{Nakajima:2014} in massive galaxies compared to 
lower-mass objects. 

To compare our result with field galaxies at the similar 
redshifts, the field SF galaxies (and some X-ray sources) 
obtained by SINS and LUCI survey \citep{Newman:2014} are 
also shown. 
The stacked spectrum `PKS1138-high' in the higher mass bin 
has a higher \oiii/\hb\ line ratio than that of `USS1558-high' 
in spite of the fact that `PKS1138-high' bin is more massive 
than that of `USS1558-high'. 
It may be caused by a significant contribution of Seyfert-type 
AGNs, and indeed `PKS1138-high' and three massive objects 
in PKS1138 are slightly offset to the AGN regime from the 
M$_\star$ vs. \oiii/\hb\ sequence which is established for 
field samples at $z>2$ (Fig. \ref{fig4}). 
However, none of the HAEs in PKS1138 are detected by 
{\it Chandra} X-ray observations \citep{Pentericci:2002}.
At this stage, we cannot clearly identify AGN contaminations.

The \oiii/\hb\ line ratio estimated from the \oiii/\ha\ 
assuming the intrinsic Balmer line ratio (\hb=\ha/2.86) 
and correcting for dust extinction, seems to work well, 
since they are consistent with the direct measurement of 
\oiii/\hb\ line ratio derived from the stacked spectra 
as shown in Fig.~\ref{fig4}. 
This method can be used as an alternative way to estimate 
the level of gaseous excitation in SF galaxies even if 
their \hb\ lines are not available. 
Note however that it tends to overestimate the \oiii/\hb\ 
line ratios for low-mass galaxies ($<$10$^{10}$\msun), 
because the dust correction using UV/\ha\ does not work 
well for young galaxies due to its sensitivity to the 
amount of hot massive stars \citep{Wuyts:2013}. 
Indeed we find that the \oiii/\hb\ line ratios estimated 
using this calibration are higher by 0.2--0.3 dex 
compared to their \oiii/\hb\ line ratios directly 
calculated by follow-up spectroscopy with the Multi-Object 
Spectrometer for InfraRed Exploration (MOSFIRE; 
\citealt{McLean:2012}) on Keck-1 10m telescope 
(Shimakawa et al. in prep.). 
Even considering this factor, however, we can say that 
the protocluster galaxies also show high gaseous 
excitation levels, same as those of field galaxies at 
similar redshifts on average. 

\subsection{BPT diagram}

The BPT diagram is the most traditional diagnostic to 
investigate the physical states of galaxies and identify 
AGN contributions \citep{Baldwin:1981}. 
It can discriminate Seyfert AGNs from \hii\ region-like 
SF galaxies based on \oiii$\lambda5007$\AA, 
\oi$\lambda6300$\AA, \nii$\lambda6583$\AA, 
\sii$\lambda6716+6731$\AA, and hydrogen Balmer series 
lines, such as \ha$\lambda6563$\AA and \hb$\lambda4861$\AA. 
In principle, low-ionization nuclear emission-line regions 
(LINERs) are also classified as AGNs 
on the BPT diagram. 
However, our previous NB imaging originally selected HAEs 
with EW$_\mathrm{H\alpha}>20$\AA\ \citep{Koyama:2013a, 
Hayashi:2012}, and the contamination of LINERs should be 
negligible since their EW$_\mathrm{H\alpha}$ is small 
($<6$\AA) as they are dominated by old (red) stellar 
continuum \citep{Cid:2011}. 
In this work, we use the \nii/\ha\ versus \oiii/\hb\ 
diagnostic which is more favorably used than other line 
diagnostics such as those using \oi\ or \sii\ since these 
lines are too weak to be detected for high redshift 
galaxies in most cases.

The theoretical predictions on the BPT diagram has been 
improved by many authors (e.g. \citealt{Veilleux:1987, 
Kewley:2001, Kauffmann:2003, Kewley:2006}).
However, applicability of the BPT diagram is disputable 
for separating SF galaxies from AGNs. 
High-$z$ galaxies, especially those at $z>2$, are 
systematically offset from the tight sequence (`abundance 
sequence'), as \oiii/\hb\ and/or \nii/\ha\ line ratios 
of $z>2$ SF galaxies tend to be much higher than those 
of low-$z$ \citep{Erb:2006, Newman:2014, Kewley:2013a, 
Masters:2014, Steidel:2014}. 
Also, the AGN branch defined at low-$z$ on the BPT diagram
seems to be shifted to the upper-left direction of the 
diagram at high-$z$ \citep{Kewley:2013a, Kewley:2013b}. 
\citet{Kewley:2013a} have newly defined the demarcation 
curve semi-empirically depending on redshift 
($0<z<2.5$) on the diagram called "cosmic BPT diagram". 
They adjust several physical parameters such as lower 
gaseous metallicity,
higher ionization parameter, and harder radiation field
to reproduce the distant SF galaxies (and AGNs).

%%%%%%%%%%%%%
\begin{figure}
\centering
\includegraphics[width=80mm]{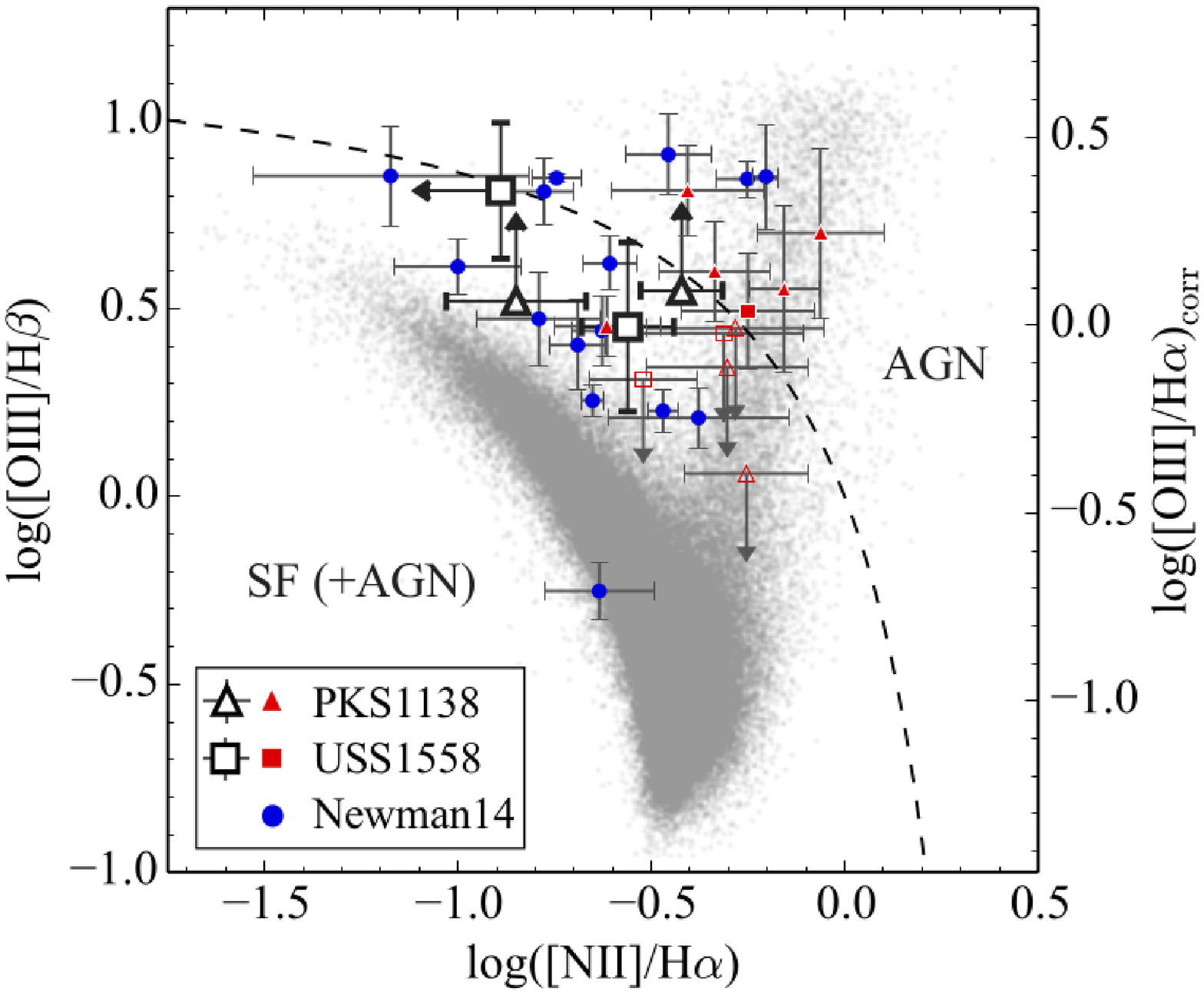}
\caption{\nii/\ha\ versus \oiii/\hb\ diagram (BPT diagram). 
White triangles and squares indicate the stacked spectra of 
HAEs in PKS1138 and USS1558 protoclusters, respectively.
Red triangles, squares, and blue circles show the individual 
HAEs in PKS1138, USS1558, and SINS \&\ LUCI data (field) 
at $z$=2.1--2.5 \citep{Newman:2014}, respectively. 
We plot only \nii\ detected objects for individual galaxies 
in our sample. 
\oiii/\hb\ line ratios of individual objects in PKS1138 and 
USS1558 are estimated with the dust extinction prescription 
of \citet{Wuyts:2013} and assuming the intrinsic \ha/\hb\ 
flux ratio of 2.86 (Case-B).
Red open symbols show the 2$\sigma$ upper limits for 
\oiii-undetected ($<2\sigma$) objects. 
\hb\ absorption is not considered for the \citet{Newman:2014} 
sample. 
Grey dots represent the SDSS galaxies with 
EW$_\mathrm{H\alpha}$$>$20\AA\ \citep{Abazajian:2009}. 
Error-bars indicate flux errors of 1 sigma of each sample. 
}
\label{fig5}
\end{figure}
%%%%%%%%%%%%%

Figure \ref{fig5} shows the BPT diagram with our data 
points from the stacked spectra and the individual 
measurements for the objects whose \nii\ line is detected 
above 2$\sigma$. 
For comparison with lower-$z$ galaxies and the field 
galaxies at $z$=2.1--2.5, we also plot other data taken 
from the literature (\citealt{Abazajian:2009, Newman:2014}, 
respectively). 
At $z>2$, the chemical abundance sequence has higher 
\oiii/\hb\ ratios on average than that of local SF galaxies 
as seen in Fig.~\ref{fig5}. 
We are not able to detect a \hb\ line for individual 
objects in the current MOIRCS spectra, and we thus estimate 
the \oiii/\hb\ line ratio from the \oiii/\ha\ ratio with 
taking into account the dust extinction and Balmer line 
ratio (\hb=\ha/2.86).
For the stacked data, we measure directly \hb/\oiii\ 
ratios whose extinctions are corrected by Balmer decrement. 

Based on the BPT diagram, we newly find five AGN 
candidates in our sample. 
They have large \nii/\ha\ and \oiii/\hb\ (from \oiii/\ha) 
line ratios above the criteria defined by 
\citet{Kewley:2013a} (cosmic BPT diagram at $z=2.5$). 
They have higher \nii/\ha\ and \oiii/\hb\ line ratios as 
is also seen in the MEx diagram.
The AGN candidates in PKS1138 are not detected in X-ray 
to a limit of $0.8\times10^{-15}$ erg/s/cm$^2$ in the 
soft band \citep{Pentericci:2002}, however, which may 
indicate that they are highly obscured AGNs. 
We exclude these AGN-like HAEs demarcated by this diagram 
from the stacked spectra hereafter, which can provide us 
with the information about \hii\ region like properties 
of the protocluster galaxies more directly. 
We caution that our spectra are unable to constrain
below log(\nii/\ha)$\lesssim$-0.5. 

\subsection{Mass--metallicity relation}

The gaseous metallicity is one of the crucial physical 
quantities to trace the chemical enrichment histories 
hence star formation histories of galaxies. 
There is a well established relationship between 
stellar-mass (or luminosity) and metallicity known as 
the mass--metallicity relation (M--Z relation) 
\citep{Tremonti:2004},
and it reflects information of gas accretion, chemical 
nucleosynthesis and outflows of galaxies. 
While the metallicity increases with age, stellar-mass 
grows not only with age but also through galaxy-galaxy 
mergers. 
Galaxies with larger gas fractions also tend to have 
lower metallicities \citep{Mannucci:2010, Tadaki:2013}. 
In dense regions, we can test whether galaxies have 
higher metallicity (are more evolved) due to 
``accelerated" galaxy formation. 
Moreover, the gas that is ejected from a galaxy by a 
galactic wind may fall back on to the host galaxy on 
a shorter timescale in denser environments 
\citep{Dave:2011b}. 
In fact, \citet{Kulas:2013} show an offset of gaseous 
metallicity in the low-mass protocluster galaxies 
(M$_\star\lesssim10^{11}$\msun) at $z=2.3$. 

Gaseous metallicity can be estimated in many ways such as 
N2 index \citep{Pettini:2004}, R23 \citep{Mcgaugh:1991, 
Kobulnicky:2004}, and a direct T$_e$ method \citep{Izotov:2006}. 
The absolute values of metallicities depend on the method to 
take, and thus conversion factors are required to compare 
metallicities among different studies \citep{Kewley:2008}. 
Several authors have compared various metallicity calibrations, 
and obtained consistent metallicities based on different line ratios 
\citep{Nagao:2006, Maiolino:2008, Troncoso:2014}. 
However, the gaseous metallicity strongly correlates with 
the ionization parameter except for some specific line 
ratios such as \nii/\oii\ \citep{Kewley:2002}. 
Also, if ISM conditions of SF galaxies change with the 
cosmic time as suggested by \citet{Shirazi:2014}, it makes 
such line diagnostics difficult and the comparison of 
gaseous metallicities uncertain between galaxies at 
different redshifts. 
\citet{Nakajima:2013, Nakajima:2014} have claimed that 
distant galaxies at $z>2$ have higher ionization parameter 
by 4 times than local galaxies, which provides a different 
metallicity for a given R23 value. 
Any reliable diagnostic for high redshift galaxies has not 
been established so far, and it should be noted that this 
work discusses only ``relative'' values of gaseous 
metallicities at similar redshift (z$\sim$2). 

We estimate a gaseous metallicity using the N2 index 
measured from the stacked spectrum after remaining AGN 
candidates (\S3.3). 
The N2 index is defined by the following expression 
\citep{Pettini:2004}, 
\begin{equation}
12 + \log(\mathrm{O/H}) = 8.90 +0.57\ \mathrm{N2}.
\label{eq4}
\end{equation}
This diagnostic is quite useful and commonly used by past 
studies (e.g. \citealt{Erb:2006}) since it requires only 
\nii$\lambda$6583\AA\ and \ha$\lambda$6563\AA\ lines.
Because of the close proximity of the two lines, this 
index is free from uncertainties in flux calibration or 
dust extinction. 
The N2 index has been empirically calibrated with local 
galaxies, and it gives an unique solution of metallicity 
as it monotonically increases with metallicity unlike the 
R23 index which often gives two solutions. 
It is thus an useful diagnostic for a relative comparison 
of gaseous metallicities among galaxies.

Although \citet{Pettini:2004} give a better fit to the data 
by a third-order polynomial function as also noted by 
\citet{Nagao:2006}, we here adopt eq. (\ref{eq4}) in order 
to compare our results directly with previous work \citep{Erb:2006}. 
The \nii\ line is weak for individual galaxies in our 
sample, but it is detected at more than 2$\sigma$ 
significance in the stacked spectra of PKS1138 (of all mass 
bins) and USS1558 (only at the high mass bin). 

%%%%%%%%%%%%%
\begin{figure}
\centering
\includegraphics[width=80mm]{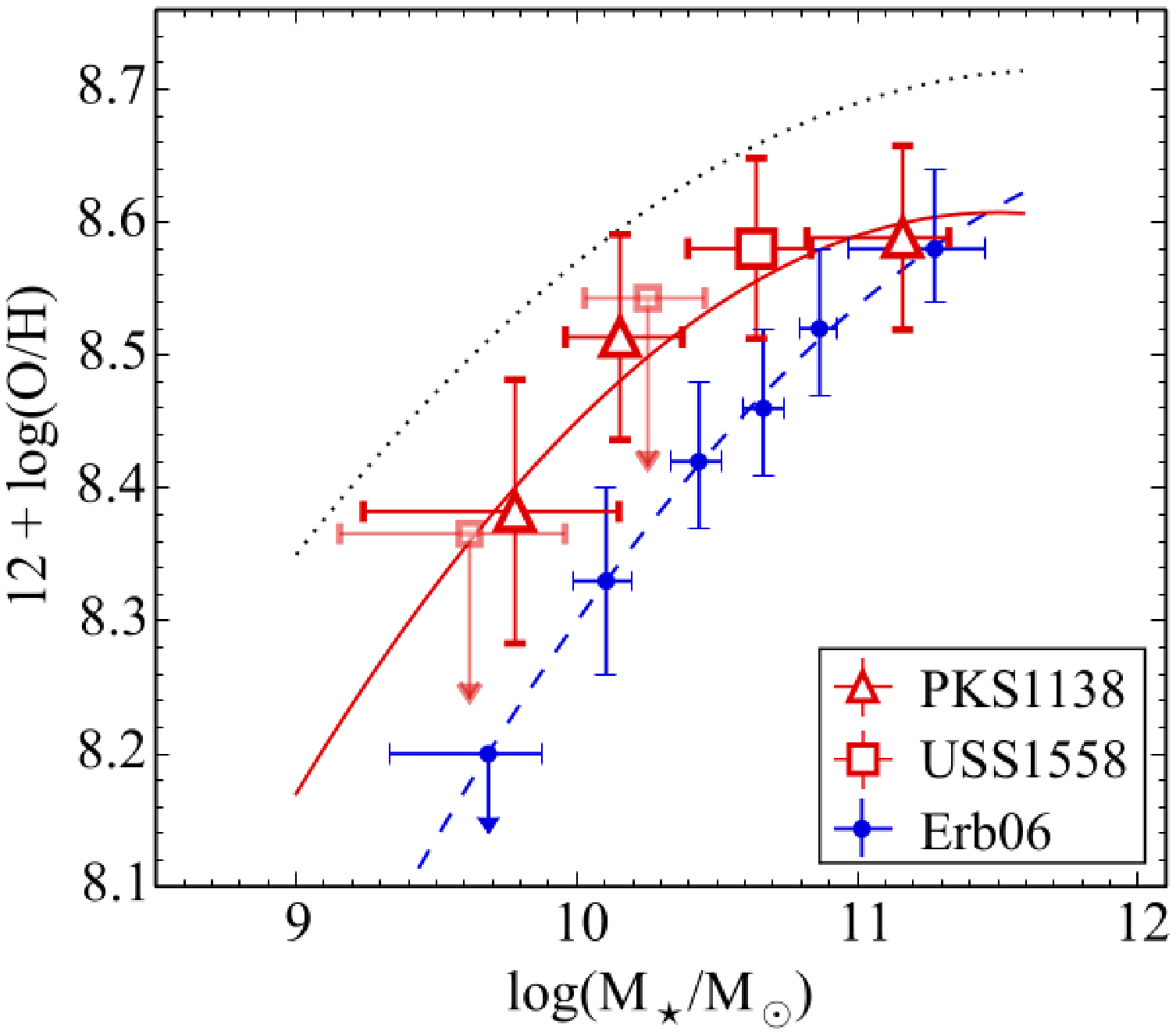}
\caption{Stellar-mass versus gaseous metallicity diagram. 
We here estimate the metallicities based on the \nii/\ha\ line 
ratios using the empirical calibration \citep{Pettini:2004}. 
Open triangles and squares indicate the stacked spectra of in 
PKS1138 and USS1558, respectively. 
Here is the revised metallicity corrected for AGN contamination 
based on the X-ray observation \citep{Pentericci:2002} 
and the cosmic BPT diagram \citep{Kewley:2013a}. 
We employ the 2$\sigma$ upper limit for \nii-undetected 
($<2\sigma$) objects. 
Blue circles indicate the metallicities measured from the 
stacked spectra of UV-selected field galaxies at $z=2.2$ 
\citep{Erb:2006}. 
The red and blue curves are the best-fit curves for each result. 
Error-bars indicate flux errors of 1 sigma in the respective 
samples. 
Dotted line represents the median of M--Z relation for SDSS 
galaxies at $z\sim0.1$ with EW$_\mathrm{H\alpha}$$>$20\AA\ 
\citep{Abazajian:2009}. 
}
\label{fig6}
\end{figure}
%%%%%%%%%%%%%

Figure \ref{fig6} shows stellar masses versus gaseous 
metallicities 
of SDSS galaxies, field galaxies at $z=2.2$ \citep{Erb:2006}, 
and our protocluster galaxies measured from the stacked 
spectra (\S2.2). 
The metallicities of the galaxies in the PKS1138 protocluster 
at $z=2.2$ (red triangles) are systematically shifted upward 
by $\gtrsim$0.1 dex from the best-fit curve of the $z=2.2$ 
field galaxies shown by the blue dashed curve \citep{Erb:2006}. 
The data point of `USS1558-high' is also located on the same 
sequence of PKS1138, which suggests that the SF galaxies in 
both protoclusters have similar chemical abundance for 
a given stellar-mass. 
It should be noted that the stacked spectra, especially 
`PKS1138-high' may be contaminated by AGNs as discussed 
in \S3.3, which raises the \nii/\ha\ line ratio. 
We have therefore excluded those AGN candidates demarcated 
by the cosmic BPT diagram \citep{Kewley:2013a} from the 
stacked spectra. 
It has lowered the data point by $\sim$0.1 dex with respect 
to the original point of `PKS1138-high'. 
As a result, we have confirmed the clear excess in gaseous 
metallicities in protocluster galaxies at the mass range of 
$\lesssim10^{11}$ \msun. 
The recent result on another protocluster at z=2.3 by 
\citet{Kulas:2013} with MOSFIRE on Keck, is consistent with 
our result as it also reports a higher metallicity in the 
dense environment than in the field. 

We approximately fit the mass--metallicity relation to our 
sample by a second-order polynomial function as follows, 
\begin{equation}
12 + \log(\mathrm{O/H}) = 8.46 + 0.21 x - 0.07 x^2,
\label{eq5}
\end{equation}
where $x\equiv\log$(M$_\star$/M$_\odot$)$-10$. 
We will use this calibration to estimate the contamination 
from \nii\ line flux and subtract it from the NB flux 
(=\ha$+$\nii) to obtain \ha\ flux (\S3.1).

%------------------------------------------------

%%%%%%%%%%%%
%%   Discussion  %%
%%%%%%%%%%%%
\section{Discussion}
Protoclusters are the ideal laboratories to study how the 
strong environmental dependence in galaxy properties seen 
today was initially set up. 
Strong line ratios such as \nii/\ha\ can provide information 
on the formation and evolution of galaxies accelerated 
environments.
In this section, various causal connections, implications, 
and speculations regarding the forming process of high-$z$ 
galaxies and its environmental dependence will be discussed 
based on the results of this work.

%------------------------------------------------
\subsection{High ionization states of SF galaxies at $\mathbf{z>2}$}
The BPT and MEx diagrams show that the SF galaxies in the 
protoclusters at $z>2$ have high ionization states even for massive 
objects. 
Such a high excitation level seen in high-$z$ galaxies have 
been reported by many recent studies in the general fields
\citep{Erb:2006, Kewley:2013a, Newman:2014, Holden:2014, 
Masters:2014, Steidel:2014}. 
Line-luminous galaxies actually tend to be located at the 
upper side of the abundance sequence on the BPT diagram 
\citep{Juneau:2014}.
It seems unlikely that this is due to an observational bias 
where we may be selectively sampling galaxies with higher 
\oiii/\hb\ ratios 
with higher completeness in our $z>2$ sample.
For example \citet{Kewley:2013a} have confirmed that their 
sample of $z>2$ galaxies do not suffer from such a bias 
originated from the sensitivity limit.
%%, i.e., the lower limit of the \oiii/\hb\ line ratio as a function of redshift. 

What is the major intrinsic cause of the high excitation 
level (higher \oiii/\hb\ ratio for a given \nii/\ha) at high 
redshifts? 
It can be attributed to different ISM conditions (ionization 
parameter, hardness of radiation field, and electron density) 
of high redshift SF galaxies compared to those of lower 
redshift counterparts as discussed in \citet{Kewley:2013b, 
Shirazi:2014}. 
In this work, we discuss more quantitatively the offset of 
the abundance sequence at $z>2$ based on the gaseous 
metallicities and the specific star-formation rates (sSFR) 
of \ha-selected galaxies.
sSFR is also an important parameter that controls the 
ionization states \citep{Shimakawa:2014c}, and indeed 
\citet{Brinchmann:2008a} have found that the SDSS 
galaxies with higher \ha\ equivalent width (almost equivalent 
to higher sSFR) are slightly elevated ($\sim$0.3 dex by 
a 3 times larger equivalent width) from the average 
sequence of the full-sample of SF galaxies.
Furthermore, sSFR increases dramatically as we go closer 
to $z\sim2$ \citep{Daddi:2007a, Whitaker:2012} while gaseous 
metallicity decreases \citep{Erb:2006, Maiolino:2008, Troncoso:2014}.
According to \citet{Whitaker:2012}, sSFR of $z=2$ SF galaxies 
at log(M$_\star$/\msun)=10.5 is $\sim$33 times larger than 
that of $z=0$ galaxies. 
There are few such galaxies with extremely high sSFRs 
in the low-$z$ universe. 
It is thus considered that such significantly high sSFRs at 
$z\sim2$ are probably responsible for their very high 
ionization states. 

%%%%%%%%%%%%%
\begin{figure}
\centering
\includegraphics[width=82mm]{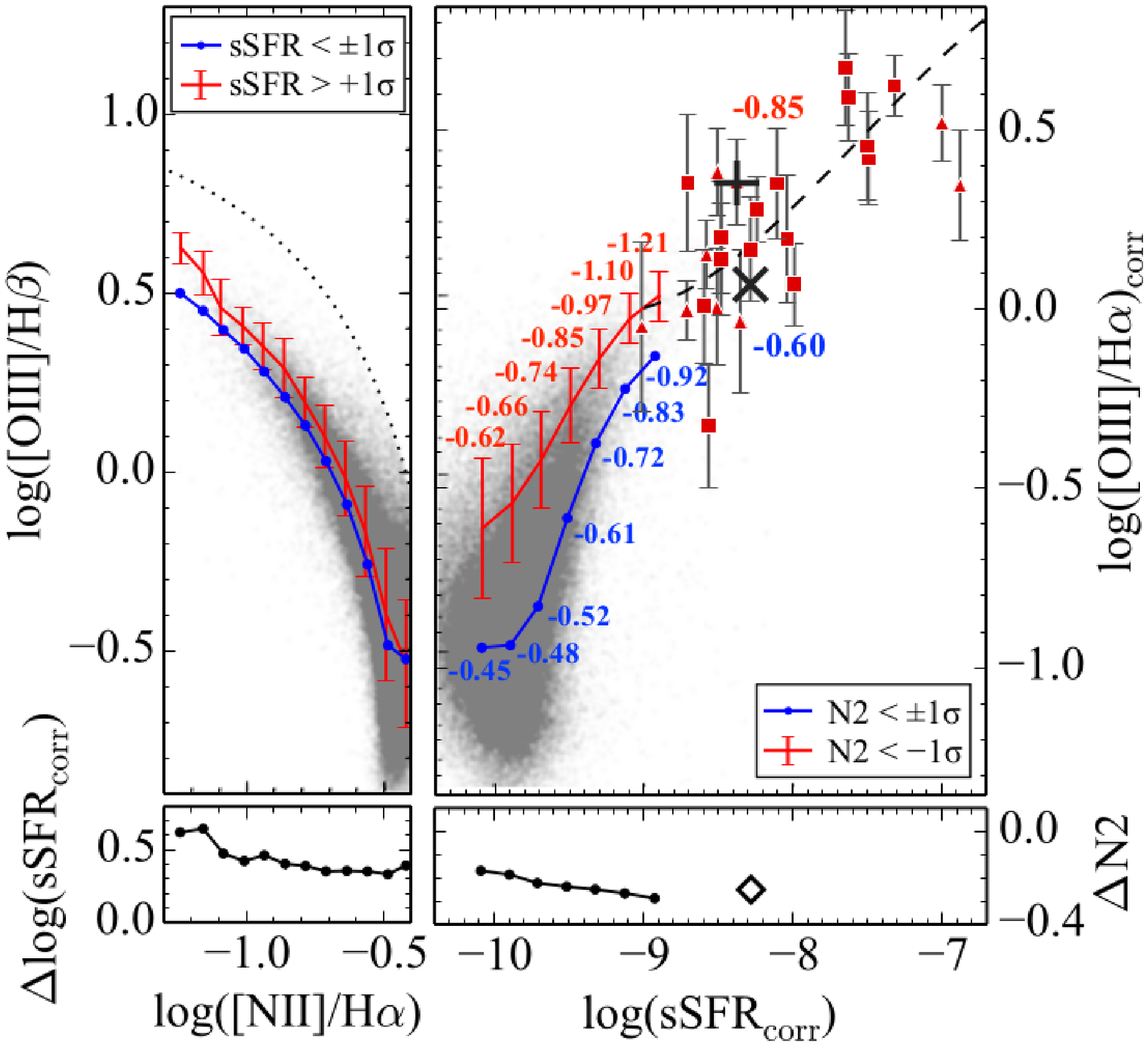}
\caption{(a) Left: \nii/\ha\ versus \oiii/\hb\ diagram 
(BPT diagram).
Grey dots show the SDSS galaxies at 0.04$\le$$z$$\le$0.1 
(\citealt{Abazajian:2009}). 
The blue curve shows the average values of the \oiii/\hb\ 
ratios as a function of N2 index (=\nii/\ha\ ratio or 
metallicity) for the SDSS galaxies whose sSFRs are within 
$\pm$1$\sigma$ scatter (nearly averaged values) for a given 
\nii/\ha\ ratio, while the red curve shows the average values 
of those whose sSFRs are higher than $+$1$\sigma$ values. 
The error-bars on the red curve show $\pm$1$\sigma$ scatter 
of the \oiii/\hb\ ratios. 
The difference in log(sSFR) of the two samples is displayed 
in the lower panel. 
(b) Right: log(sSFR) versus \oiii/\hb\ diagram. 
Grey dots show the SDSS galaxies (same as those in the left 
panel). 
The blue curve shows the average values of \oiii/\hb\ ratios 
as a function of sSFR for the SDSS galaxies whose N2 indices 
(hence metallicity) are within $\pm$1$\sigma$ scatter (nearly 
averaged values) for a given sSFR, while the red curve shows 
the average values of those whose N2 indices are smaller 
than $-$1$\sigma$ values. 
The error-bars on the red curve show $\pm$1$\sigma$ scatter 
of the \oiii/\hb\ ratios.
The difference in N2 values of the two samples is displayed 
in the lower panel. 
Filled triangles and squares indicate our measurements for 
the individual HAEs in PKS1138 and USS1558 protoclusters, 
respectively. 
We note that the individual \oiii/\hb\ line ratios are 
actually derived from the \oiii/\ha\ ratios, by estimating 
the dust extinction based on the UV/\ha\ ratios (which we 
know are consistent with the Balmer decrement measurements 
from the stacked spectra (see \S3.2). 
The ``$+$" and ``$\times$" symbols represent the median 
values for the two samples of HAEs separated by the dashed 
line which is the best-fit line to our sample.
The averaged N2 values are indicated by small numbers at each 
bin, including the above two HAE samples (measured from the 
stacked spectra for our data).
Note that for a given sSFR, those having higher \oiii/\hb\ ratios 
have lower N2 values, indicating that gaseous metallicities 
are also contributing to high \oiii/\hb\ ratios of the HAEs 
in the protoclusters as well as high sSFRs. 
}
\label{fig7}
\end{figure}
%%%%%%%%%%%%%

Figure \ref{fig7}a shows a BPT diagram for the SDSS galaxies.
The red and blue curves show the difference in galaxy 
distributions on this diagram between those having higher 
dust-corrected sSFR (sSFR$_\mathrm{corr}$) above the 1$\sigma$ 
scatter at a given N2 index ($\equiv$log(\nii/\ha) or gaseous 
metallicity), and those with nearly average 
sSFR$_\mathrm{corr}$ values within $\pm$1$\sigma$ scatter.
Here, we employ only the \hii\ region-like galaxies in the 
SDSS sample within a specific redshift range ($z$=0.04--0.1) 
selected by local BPT diagram \citep{Kewley:2006} and their 
sSFR are based on \citet{Brinchmann:2004}. 
We can see that the \oiii/\hb\ ratios of the galaxies with 
higher sSFRs (by $\sim0.5$ dex; see the bottom panel of 
Fig.\ 7a have systematically higher \oiii/\hb\ ratio by 
$\sim0.1$ dex for a given N2 i.e. metallicity on average).

On the other hand, Fig. \ref{fig7}b represents the 
dust-corrected sSFR (sSFR$_\mathrm{corr}$) versus \oiii/\hb\ 
line ratio and their difference in galaxy distribution on 
the diagram between those having lower N2$\equiv$log(\nii/\ha) 
(lower than 1$\sigma$ scatter at a given sSFR$_\mathrm{corr}$) 
and those having nearly averaged N2 values within 
$\pm$1$\sigma$.
Our HAEs in the protocluster are over-plotted in this 
diagram by filled triangles and squares. 
sSFR$_\mathrm{corr}$ are derived from the NB fluxes with 
the dust correction based on UV/\ha\ ratios (\S3.1). 
The \oiii/\hb\ line ratios are estimated from the \oiii/\ha\ 
line ratios using the dust correction prescription based on 
UV/\ha\ ratios (\S3.2). 
This method works well since it is consistent with the direct 
mesurements of \oiii/\hb\ ratio for the stacked spectra (\S3.2). 
Interestingly, the HAEs in the protoclusters at $z>2$ more or 
less follow the extension of the correlation between sSFR vs. 
\oiii/\hb\ for the SDSS galaxies.
Such trend is also reported in a recent paper 
\citep{Holden:2014}, but our sample covers a wider 
stellar-mass range 
(9$\lesssim$log(M$_\star$/\msun)$\lesssim$12). 
In fact, our protocluster galaxies at $z>2$ have much 
higher sSFRs (by 1--2 orders of magnitude) as well as much 
higher \oiii/\hb\ ratios compared to the SDSS galaxies. 
This suggests that sSFR is a key factor that raises 
the \oiii/\hb\ line ratios. 
Moreover, the ratios also depend on the gaseous metallicity 
in the sense that lower \nii/\ha\ ratios (N2 values) hence 
lower metallicities raise the \oiii/\hb\ line ratios for a 
given fixed sSFR.
Therefore, the significant offset of the high-z SF galaxies on 
the BPT diagram is likely to be caused by their both higher 
sSFRs and lower metallicities compared to those of local 
galaxies. 
Lower metallicities shift the galaxies to the upper left 
direction on the BPT diagram (lower \nii/\ha\ and higher 
\oiii/\hb) \citep{Kewley:2013b}.
Also, higher sSFRs raise the \oiii/\hb\ line ratios further 
at a given metallicity (N2).

Although Fig. \ref{fig7} apparently shows a smooth 
correlation between log sSFR and \oiii/\hb\ ratio, there is 
a large gap of 0.5 dex between the local SDSS galaxies and 
our samples of high-$z$ SF galaxies when we compare them 
at a fixed \nii/\ha\ line ratio. 
It suggests that some other factors may be contributing to 
the offset of the high-$z$ galaxies on the BPT diagram
such as the effects of outflows \citep{Whitaker:2014}, 
N/O enhancement by Wolf-Rayet stars \citep{Masters:2014}, 
and/or higher gas pressure in high-$z$ galaxies 
\citep{Shirazi:2014}.
We cannot resolve this problem from this result alone, 
and further details are beyond the scope of this paper.

%------------------------------------------------

\subsection{Environmental dependence of galaxy formation at $\mathbf{z>2}$} 

The PKS1138 and USS1558 are the largest class of protocluster 
at $z=2.2$ and $z=2.5$, respectively which are likely to grow 
to a massive cluster of a total mass of $\sim10^{15}$ \msun\ 
by the present-day \citep{Shimakawa:2014}. 
As such it is the ideal place to test the environmentally 
dependent galaxy formation in the early epoch. 
We have compared the physical states and the gaseous 
metallicities of HAEs in the protoclusters at $z$=2.2 and 2.5 
with those in the general fields to address the environmental 
dependence in galaxy formation.
Our data cover a wide range in stellar mass, 
i.e.\ log(M$_\star$/\msun)=9--12, which enables us to 
investigate the mass dependence of the environmental 
dependency as well. 

Generally, the outflow of chemically-enriched gas can 
reduce gaseous metallicity in galaxies, while the inflow 
of primordial gas also dilutes the metallicity 
\citep{Erb:2006,Erb:2008,Dalcanton:2007,Mannucci:2009}. 
The upward shift by 0.1--0.15 dex of the M--Z relation in 
the protoclusters with respect to that of the general field 
at $z\sim2$ indicates that the chemical evolution proceeds 
faster in the dense environment at a given stellar mass.
Recently, \citet{Kulas:2013} first noted the excess of 
gaseous metallicities of SF galaxies in the protocluster 
HS1700+643 ($z=2.3$) as compared to the general field, at 
the stellar mass range of log(M$_\star$/\msun)$\lesssim11$. 
This is consistent with our result (\S3.4). 
The reasons for such an offset of the M--Z relation in the 
protocluster at $z>2$ have five possibilities, namely, 
(1) recycling of chemically enriched gas due to higher pressure of
inter galactic medium (external or nurture effect) 
\citep{Dave:2011b, Kulas:2013}, 
(2) stripping of metal poor \hi\ gas loosely trapped in the outer
reservoirs by galaxy-galaxy interactions or ram-pressure
(external or nurture effect) 
(3) advanced stage of down-sizing galaxy evolution 
(intrinsic or nature effect) \citep{Thomas:2005}, 
(4) top-heavy IMF in young star-bursting phase, 
and (5) sample selection bias.

\begin{enumerate}
\renewcommand{\theenumi}{(\arabic{enumi})}

\item 
First of all, the environmentally dependent recycling of 
chemically enriched gas is discussed by \citet{Kulas:2013}.
They suggest that the metal enhancement seen in the 
over-dense regions can be accounted for by a fallback 
and a recycling of ejected outflowing gas due to higher 
pressures of the surrounding inter galactic medium 
(IGM) in clusters. 
Recycling of chemically enriched gas leads to the formation 
of next generation of stars and thus chemical evolution 
would be proceeded further in clusters. 
Such phenomenon is predicted by numerical hydrodynamical 
simulations by \citet{Oppenheimer:2008, Dave:2011b}.
In this scenario, the metal enhancement depends not only 
on the IGM density hence the richness of clusters, but also 
on the dynamical mass of individual galaxies.
Such differential chemical evolution between different 
environments must be more prominent for less massive 
galaxies because massive galaxies have larger potential 
wells and the gas tends to be retained and easier to be 
recycled in any case irrespective of their surrounding 
environments.
This is favored by the observations which indicate a lack 
of environmental dependence of metallicities in massive 
galaxies.

\item
Secondly, once an in-falling galaxy is incorporated into 
a common cluster halo, the gas reservoir of the galaxy 
may be stripped or truncated due to the interactions with 
cluster potential, intra-cluster gas, and/or other galaxies 
\citep{Abadi:1999, McCarthy:2008}.
This can not only expel low metallicity gas trapped in 
the outer region of the galaxy, but also terminate the fresh 
gas accretion onto the galaxy which would have otherwise 
diluted the high metallicity gas at the center in the infall 
dominated phase of galaxy formation ($z>2$). 
Therefore, we expect that this process can effectively 
increase the gaseous metallicity of galaxies in cluster 
environment, as compared to isolated galaxies. 

\item 
Thirdly, there is an intrinsic (nature) effect that the 
timescale of galaxy evolution depends on environment, 
because densest environments such as clusters of galaxies 
start off from the highest density peaks in the early 
Universe which must collapse first.
Because of this bias, galaxy formation and evolution should
proceed earlier and/or faster, so does the chemical evolution.
In this case, the environmental dependence of gaseous 
metallicity seen particularly in lower mass galaxies can 
be naturally explained by environmentally dependent galaxy 
downsizing (mass dependent timescale of galaxy formation 
and evolution, \citealt{Cowie:1996, Cattaneo:2008}), in the 
sense that the downsizing proceeds earlier in the protocluster 
regions, and hence the gas in protocluster galaxies are more
chemically enriched for a given stellar mass especially in less
massive galaxies.

One should note, however, that we are comparing 
metallicities between different environments at a given 
`stellar' mass, and not at a fixed total or `halo' mass 
including `gas' mass.
The same amount of stellar mass means that they have 
synthesized the same absolute amount of metals.
However, the gaseous metallicity (O/H) depends on gas 
content or fraction as well. 
Cluster galaxies can have the same stellar mass as the 
field galaxies but with smaller gas fraction and thus 
smaller total baryonic or halo masses, because of the 
more advanced downsizing effect.
Basically, we may be comparing between the advanced 
stage of galaxies with low halo masses in clusters with 
the less advanced stage of galaxies with high halo masses 
in the field.
Such difference is expected to be seen more prominently 
in less massive systems, since massive galaxies are 
already well evolved irrespective of their environments, 
and we do not expect to see a sizable difference in the 
evolutionary stage hence gaseous metallicity for massive galaxies.

This scenario, however, has a contradiction. 
Because the advanced evolutionary stage should mean 
smaller gas fraction for cluster galaxies for a given 
stellar mass, we would expect them to have lower specific 
sSFRs compared to those of field galaxies. 
This is not consistent with the observational results of 
\citet{Koyama:2013a, Koyama:2013b} who show the 
environmental independence of sSFRs.

\item 
A top heavy IMF for cluster galaxies can also explain the 
metallicity excess since it increases the intrinsic chemical 
yield. 
The top-heavy IMF is actually preferred to account for high 
($\alpha$/Fe) ratios of massive early-type galaxies (e.g. 
\citealt{Baugh:2005, Nagashima:2005a, Nagashima:2005b}).
However, the metallicity difference is seen only in low mass 
galaxies, and so this scenario is not favored.

\item 
Finally, different sample selections may be an issue between 
this work (protoclusters) and \citet{Erb:2006} (general field). 
Our NB selection of SF galaxies based on \ha\ emission line 
can cover a broader range of SF galaxies in terms of 
stellar-mass. 
Also the dust obscuration effect can be minimized.
On the contrary, the UV-selected SF galaxies by 
\citet{Erb:2006} tend to be biased to less dusty and 
relatively young populations (see also \citealt{Steidel:2004}).
This may result in an apparent difference in the M--Z relation 
between the two samples as discussed by \citet{Stott:2013} at 
$z$=0.8--1.5.
These two samples correspond to different stages of galaxy 
evolution in the sense that \ha\ selected galaxies are in 
a more advanced stage hence could be chemically more enriched 
at a given stellar mass.
However, there is no environmental dependence between field 
and cluster galaxies at $z\sim2$ as reported by 
\citet{Koyama:2013a, Koyama:2013b}. 
Moreover, the offset we see in the M--Z relation for the 
protocluster galaxies (0.1--0.15 dex) cannot be accounted
for by the sampling bias alone that \citet{Stott:2013} claim.
In fact, for the SDSS galaxies, the amount of dust
extinction is largely determined by the stellar mass,
and at the fixed stellar mass we do not see any clear
dependence of the deviation from the M--Z relation
on the amount of dust extinction (see Appendix, and also
\citealt{Zahid:2014b}).
Therefore, we do not expect a strong selection bias between
the UV selected SF galaxies \citep{Erb:2006} and the 
H$\alpha$ selected samples (this work).
Furthermore, the fact that we do not see the metallicity 
offset for massive galaxies which tend to be more heavily 
obscured also suggest that the sampling bias should not 
be the main reason for the offset of the M--Z relation 
for low mass galaxies.

\end{enumerate}

%%%%%%%%%%%%%
\begin{figure}
\centering
\includegraphics[width=75mm]{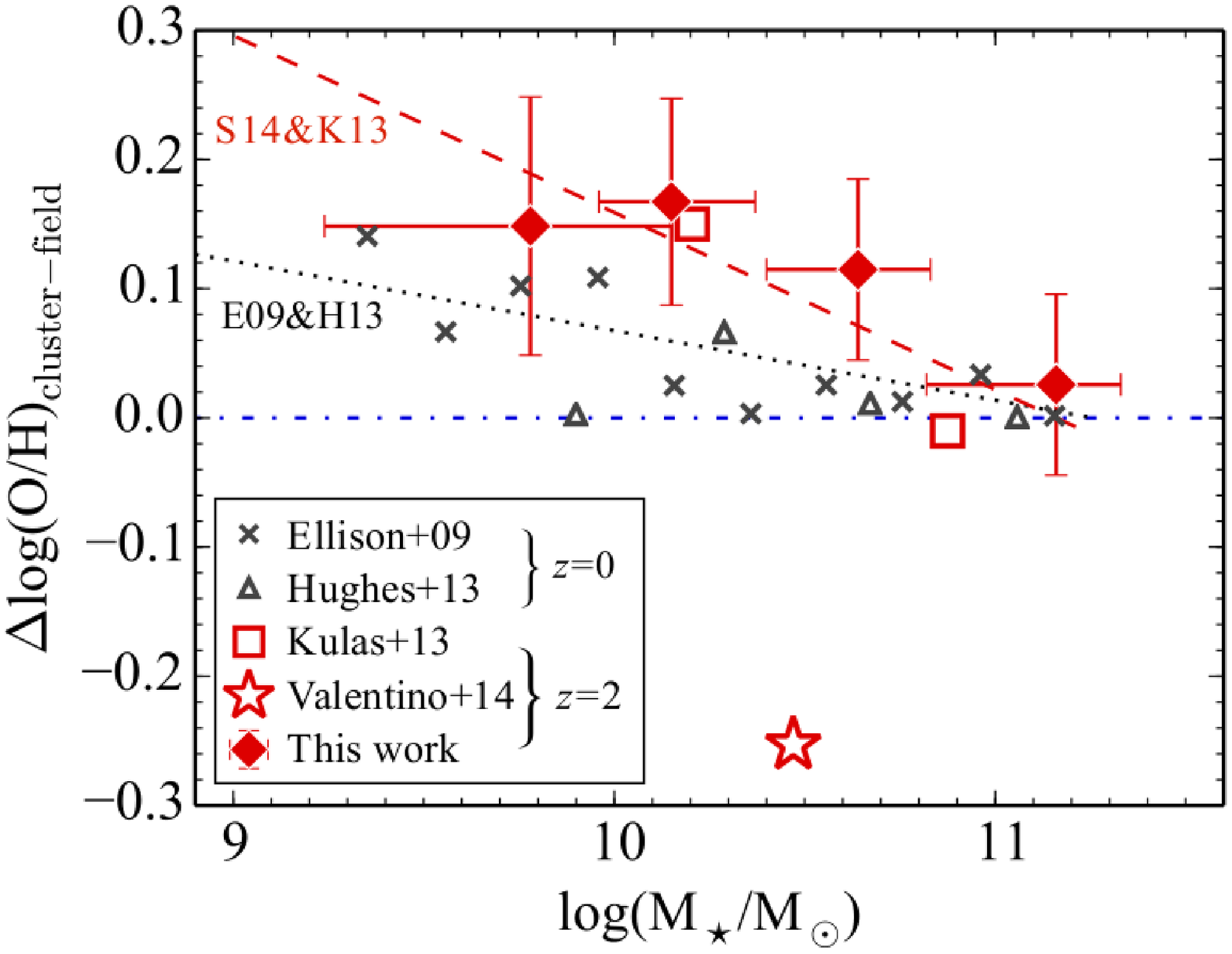}
\caption{
The offset of the averaged gaseous metallicities 12+log(O/H) 
of protocluster/cluster galaxies from those in general fields 
in the literature plotted against stellar mass. 
Red diamonds, squares, and a star indicate SF galaxies at 
$z\sim2$ in dense regions reported by this work and 
\citet{Kulas:2013, Valentino:2014}, respectively. 
Grey crosses and triangles show local SF galaxies in 
dense regions above $\Sigma>1$ Mpc$^{-2}$ based on SDSS 
sample \citep{Ellison:2009} and in Virgo cluster 
\citep{Hughes:2013}, respectively.
Dotted and dashed lines are the fitted lines where 
$\Delta$log(O/H) is set to be zero at 
log(M$_\star$/\msun)=11.16 for both local and the $z\sim2$ 
samples except for the \citet{Valentino:2014} data.}
\label{fig8}
\end{figure}
%%%%%%%%%%%%%

Following the above discussions, we suggest that the offset
in gaseous metallicity in dense regions may be driven by the 
environmental dependence of ``feeding \& feedback" processes
of galaxy formation at high redshifts. 
Cluster galaxies are formed within complicated structures such 
as clusters, surrounding groups, filaments and sheets, as 
simulated by CDM-dominated cosmological simulations 
(e.g. \citealt{Springel:2005}). 
Also, recent models predict that the cold gas accretion is the 
dominant gas supply mechanism in high redshift galaxies, and 
that such process of gas inflow depends strongly on halo mass 
and redshift \citep{Dekel:2009}. 
Moreover, other models simulate gas outflow processes
which are dependent on galaxy mass and surrounding 
environment \citep{Dave:2011b}. 
Therefore, it is natural that cluster galaxies follow 
different tracks of chemical evolution from the field ones
in their vigorous formation phase.

Figure \ref{fig8} represents the offset of gaseous metallicities 
of the cluster galaxies from those of field counterparts. 
Here, we collect $z\sim2$ samples from \citet{Kulas:2013, 
Valentino:2014} include this work, and local samples from 
\citet{Ellison:2009} (SDSS galaxies in dense region above 
$\Sigma>1$ Mpc$^{-2}$) and \citet{Hughes:2013} (Virgo cluster). 
We clearly see a metallicity enhancement by 0.15 dex around 
M$_\star>10^{10}$ \msun\ in this work, which is consistent with
the earlier result by \citet{Kulas:2013}. 
\citet{Ellison:2009} also reports such an excess of gaseous
metallicity in cluster regions at $z\sim0.1$ (see also 
\citealt{Peng:2014}). 
Interestingly, the offset of the M--Z relation in dense regions 
gets larger at $z\sim$2 (Fig. \ref{fig8}). 
This can be explained by the fact that gas transfer processes 
such as inflow and outflow are more active at $z\sim2$ 
than those at low redshift \citep{Steidel:2010, Bouche:2013, 
Yabe:2014b}. 
It should be noted however that this gives only a partial 
explanation for the environmental dependence. 
In fact, \citet{Valentino:2014} recently reports a completely
opposite trend for another overdensity region at $z=2$.

Moreover, we need to consider SF activities of these 
protocluster galaxies at the same time.
The gas stripping in protoclusters, if present, would lower the
gas fraction, and thus lead to lower SFRs compared to 
field galaxies \citep{Balogh:2000b, Kodama:2001}.
However, our samples of HAEs show similar SFRs 
irrespective of environment, and the protocluster galaxies share
the same main-sequence of field SF galaxies.
In the current studies we have to rely on the stacking analyses, 
which have lost information of any scatter among individual 
galaxies. 
Since the environmental effects are rather stochastic, a 
considerable scatter in the chemical properties would be expected. 
Therefore investigating the environmental dependence of chemical
evolution on individual galaxy basis is essential as the next step. 

%------------------------------------------------
\subsection{Fundamental metallicity relation} 

%%%%%%%%%%%%%
\begin{figure}
\centering
\includegraphics[width=75mm]{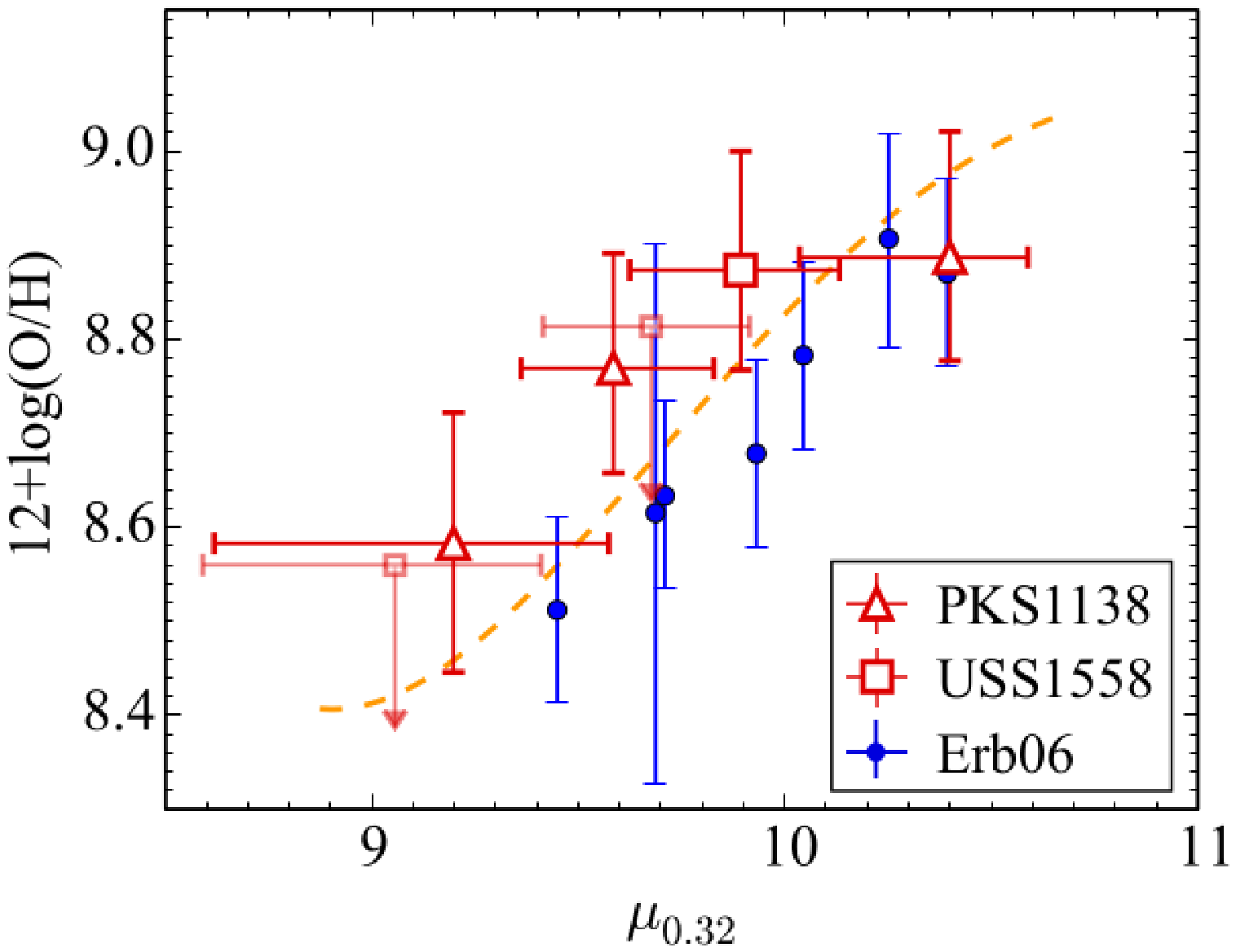}
\caption{The fundamental metallicity relation (FMR), $\mu_{0.32}$ versus 
12+log(O/H), where $\mu_{0.32}$$\equiv$log(M$_\star$)$-$0.32 log(SFR). 
Red triangles and squares indicate the protoclusters, PKS1138 at $z=2.2$ and 
USS1558 at $z=2.5$, respectively. 
Here we employ the \citet{Maiolino:2008} calibrations for 
a direct comparison with the FMR of \citet{Mannucci:2010}. 
Blue circles show the field galaxies at $z=2.2$ \citep{Erb:2006}, although their 
gaseous metallicities have been revised by \citet{Maiolino:2008}. 
The orange curve shows the FMR reported by \citet{Mannucci:2010}. }
\label{fig9}
\end{figure}
%%%%%%%%%%%%%

We finally discuss the fundamental metallicity relation (FMR) of 
the protocluster galaxies. 
\citet{Mannucci:2010} show that the scatter around the M--Z 
relation is mainly driven by the correlation between SFR and 
metallicity at a given stellar mass (see also 
\citealt{Mannucci:2011, Bothwell:2013, 
Stott:2013, Yabe:2014, Troncoso:2014}) in the sense that 
galaxies with higher SFRs tend to have lower metallicities. 
The FMR is thus expressed as 
12+log(O/H)=log(M$_\star$/\msun)$-$0.32 log(SFR) 
($\equiv\mu_{0.32}$) and it has a scatter of only 0.05 dex. 
Moreover, the form of FMR seems unchanged out to $z\sim2.5$ 
\citep{Mannucci:2010}, and probably to $z\gtrsim3$ as traced 
by Lyman break galaxies (\citealt{Nakajima:2013, Nakajima:2014}, 
but see also \citealt{Mannucci:2010, Troncoso:2014, 
Steidel:2014, Sanders:2014}). 
This means that most of the SF galaxies can evolve only on 
the FMR and that star formation and chemical enrichment in 
galaxies are well regulated through secular processes across 
the large redshift range.

An obvious extension of such studies is to investigate the 
environmental dependence of the FMR at high redshift to test 
whether the chemical enrichment history, and thus the 
regulation laws among star formation, inflow, and outflow, 
are accelerated in dense environments.
According to \citet{Magrini:2012}, any difference in the FMR 
cannot be seen for cluster galaxies at $z=1.4$.
We now show the results for higher redshift clusters at $z>2$
where massive galaxies are growing rapidly \citep{Kodama:2007}.
During this epoch of galaxy formation, galaxies may go through
violent, unstable phases such as massive gas infall 
\citep{Dekel:2009}, 
outflows by SF/AGN feedback, and galaxy-galaxy mergers 
\citep{Gottlober:2001}, which are all possibly environmentally 
dependent. 
Such environmental effects may have some impacts on chemical evolution
that are different from those driven by secular processes, 
and thus produce an environmental variation of the FMR.

Figure \ref{fig8} represents $\mu_{0.32}$ versus gaseous 
metallicities of HAEs in the protoclusters and those in the 
general field \citep{Erb:2006, Mannucci:2010}. 
We have revised the gaseous metallicities based on the 
\citet{Maiolino:2008} prescription (\nii/\ha\ ratio) instead 
of the \citet{Pettini:2004} method, in order to directly compare 
the metallicities of our sample with those by 
\citep{Mannucci:2010} with the same method. 
\begin{equation}
\mathrm{N2} = -0.773 +1.236 x -0.281 x^2 -0.720 x^3 -0.333 x^4
\end{equation}
Here, $x$ is 12+log(O/H). 
We adopt the median values of stellar mass and SFR at 
each bin, and the error-bars indicate $1\sigma$ scatters 
of the stacked spectra. 
Our protocluster data are located above the fundamental 
plane (orange curve) by 0.10--0.15 dex at 
$\mu_{0.32}\lesssim10$ although the plots except 
`PKS1138-mid' are still consistent within the errors. 
The slight ($\sim$0.10--0.15) offset of the FMR of the 
protocluster galaxies is caused by the 
metal enhancement at the stellar mass smaller than 
$10^{11}$\msun. 
It should be noted that the extinction correction is different 
between this work and \citet{Erb:2006}, and our plots would 
be shifted rightward by $\lesssim0.1$ dex if we adopt the 
same dust correction prescription by \citet{Erb:2006} (see 
also \citealt{Erb:2006b}). 
Because of too large errorbars, however, we cannot 
constrain the FMR of the protocluster galaxies. 

Moreover, there are some additional caveats for the investigation
of the fundamental metallicity relation. 
Firstly, the high-$z$ SF galaxies have higher ionization 
parameters and higher electron densities, which may lead us to
have a systematic error in gaseous metallicity measurement
\citep{Nakajima:2014}. 
For example, \citet{Steidel:2014} show that the abundance 
measurement of a high-$z$ galaxy based on N2 index tend to be
an overestimation compared to the direct measurement based on the
T$_e$ method \citep{Izotov:2006}.
Secondly, we see no clear dependence of the M--Z relation on 
SFR in the recent works \citep{Steidel:2014, Sanders:2014}. 
Therefore it may be too early to discuss the environmental 
dependence of the fundamental metallicity relation yet. 

%------------------------------------------------
\section{Conclusions}

%%%%%%%%%%%%%
\begin{table*}
\begin{center}
 \caption{Final catalogue of the stacked spectra of HAEs in PKS1138 and USS1558 separated into two stellar 
 mass bins (low and high) and the same stellar-mass bin for comparing with each protocluster (mid). 
 We here exclude AGN candidates (see \S3.3) from the sample. 
 (1) ID categorized by stacking method, (2) the median stellar-mass, 
 (3)--(4) line ratios of median stacked spectra, 
 (5) gaseous metallicities derived from PP04 (N2 index), and (6) dust extinction of \ha\ calculated by using Balmer decrement. 
 All errors are derived from 1$\sigma$ scatters of respective emission line fluxes. }
 \begin{tabular}{@{}cccccc@{}}
\hline
  ID & log(M$_\star$/M$_\odot$) & log(\nii/\ha) & log(\oiii/\hb) & 12$+$log(O/H) & A$_\mathrm{H\alpha}$ \\
  (1) & (2) & (3) & (4) & (5) & (6) \\
\hline
PKS1138-low & $9.78_{-0.54}^{+0.37}$ & $-0.91\pm0.17$ & $>0.42$ & $8.38\pm0.10$ & $>2.06$ \\
PKS1138-high & $11.16_{-0.34}^{+0.17}$ & $-0.55\pm0.12$ & $>0.19$ & $8.59\pm0.07$ & $>1.96$ \\
\hline
USS1558-low & $9.62_{-0.46}^{+0.34}$ & $<-0.94$ & $0.87\pm0.18$ & $<8.37$ & $1.28\pm1.15$ \\
USS1558-high & $10.64_{-0.24}^{+0.19}$ & $-0.56\pm0.12$ & $0.45\pm0.22$ & $8.58\pm0.07$ & $1.72\pm1.41$ \\
\hline
PKS1138-mid & $10.15_{-0.19}^{+0.22}$ & $-0.68\pm0.14$ & $>0.34$ & $8.51\pm0.08$ & $>0.78$ \\
USS1558-mid & $10.25_{-0.23}^{+0.20}$ & $<-0.63$ & $>0.51$ & $<8.54$ & $>1.10$ \\
\hline
\end{tabular}
\end{center}
\label{tab3}
\end{table*}
%%%%%%%%%%%%%

This paper reports the results on the physical properties 
of HAEs in two protoclusters, PKS1138 at $z=2.2$ and USS1558 
at $z=2.5$, based on the NIR spectroscopy with MOIRCS on Subaru. 
We investigate the environmental dependence of the physical 
properties of SF galaxies to test whether how the SF activity 
is intrinsically biased in the dense environments, and how they 
are affected by their surroundings at the peak epoch of galaxy 
formation.
%%, which eventually establish the tight environmental 
%%dependency of galaxy properties seen in the present-day Universe.

\begin{description}
\item[--] 
We measure the strength of dust attenuation of HAEs using 
Balmer decrement (\ha/\hb) from the stacked spectra and 
SFR(\ha)/SFR(UV) ratios of individual galaxies. 
We then investigate their physical properties with an aid of 
the MEx diagram \citep{Juneau:2011, Juneau:2014} and find 
that the ionization states of the protocluster galaxies at 
$z>2$ is much higher than those of the field galaxies
by an order magnitude based on their \oiii/\hb\ 
or \oiii/\ha\ line ratios. 
This is consistent with many other recent works on SF galaxies 
at $z>2$ in general fields \citep{Erb:2006, Kewley:2013a, 
Kewley:2013b, Troncoso:2014, Newman:2014, Masters:2014, 
Steidel:2014, Shapley:2014, Coil:2014}. 
Such a high gaseous excitation may be primarily driven by 
their low metallicities and very high sSFRs as shown in 
Figure \ref{fig7}. 
However, this is not a comprehensive explanation for the
high ionization of the $z\sim2$ galaxies, and some other effects
would be needed as well \citep{Shirazi:2014, Masters:2014, Wuyts:2014}.

\item[--] 
The history of chemical enrichment provides us with 
information on the physical processes that are occurring 
during the epoch of galaxy formation. 
We explore the environmental dependency of the gaseous 
metallicity using the N2 index \citep{Pettini:2004}. 
We exclude the AGN contamination by using the BPT diagram 
(\nii/\ha\ vs. \oiii/\ha) and X-ray data. 
We find that the protocluster galaxies of low masses 
($\lesssim10^{11}$ \msun) tend to be more chemically enriched than 
the field counterparts at $z=2.2$ \citep{Erb:2006}. 
In order to account for the environmental dependence in the
M--Z relation, we discuss five possible scenarios, including 
recycling of chemically enriched gas and/or stripping of 
metal poor gas in the reservoir.
Among them, we suggest that the metal enhancement is most likely
caused by the environmental effects on feeding \& feedback processes
during galaxy formation. 
The higher offset of the M--Z relation observed in dense environments
at $z\sim2$ with respect to the local relation can be explained by
very active inflow and outflow processes at that epoch
\citep{Steidel:2010, Bouche:2013, Yabe:2014b}. 
This scenario is also supported by some theoretical works
(e.g. \citealt{Dave:2011b}). 
However, we have not yet reached to any fully consistent picture from
the chemical evolution alone which can also account for the SF
activities in protocluster galaxies.
Further investigations, both observationally and theoretically, 
are clearly needed.

\end{description}

The largest limitation of our current study is that we cannot 
derive gaseous metallicities for individual HAEs, and we have 
to rely on the stacking spectral analysis which gives
us only averaged metallicities in each stellar mass bins. 
This cannot resolve the ``heterogeneity" of chemical evolution in
SF galaxies in dense environments in their early formation phase.
We need deeper spectroscopy which allows us to detect multiple 
emission lines in each galaxy and measure its 
metallicity with the ISM condition also taken into account.
More specifically, ionization parameter can be estimated by 
\oiii/\oii\ or \neiii/\oii\ ratios \citep{Levesque:2014}, and 
electron densities can be derived directly from 
\oii$\lambda$3729/3726 or \sii$\lambda$6716/6731 ratios 
\citep{Puech:2006, Krabbe:2014, Shimakawa:2014c}.
Such observations have just become possible by the advanced 
NIR multi-object spectrographs such as MOSFIRE on Keck.

%------------------------------------------------

\section*{Acknowledgments}

This paper is based on the data collected at Subaru Telescope, which is operated by the
National Astronomical Observatory of Japan.
We acknowledge Dr. K. Aoki and Dr. K. Yabe at Subaru Telescope 
for useful discussion. 
This work is financially supported in part by a Grant-in-Aid 
for the Scientific Research (Nos.\, 21340045 and 24244015) 
by the Japanese Ministry of Education, Culture, Sports,
Science and Technology. 
We thank the referee for acknowledgements and kind help 
to improve the statement. 

%%%%%%%%%%

\bibliographystyle{mn2e}
\bibliography{bibtex_library}

\begin{thebibliography}{132}
\expandafter\ifx\csname natexlab\endcsname\relax\def\natexlab#1{#1}\fi

\bibitem[{{Abadi}, {Moore} \& {Bower}(1999){Abadi}, {Moore}, \&
  {Bower}}]{Abadi:1999}
{Abadi} M.~G., {Moore} B., {Bower} R.~G., 1999, \mnras, 308, 947

\bibitem[{{Abazajian} {et~al}\mbox{.}(2009){Abazajian}, {Adelman-McCarthy},
  {Ag{\"u}eros}, {Allam}, {Allende Prieto}, {An}, {Anderson}, {Anderson}, \&
  et~al.}]{Abazajian:2009}
{Abazajian} K.~N. {et~al.}, 2009, \apjs, 182, 543

\bibitem[{{Adelberger} {et~al}\mbox{.}(2004){Adelberger}, {Steidel}, {Shapley},
  {Hunt}, {Erb}, {Reddy}, \& {Pettini}}]{Adelberger:2004}
{Adelberger} K.~L., {Steidel} C.~C., {Shapley} A.~E., {Hunt} M.~P., {Erb}
  D.~K., {Reddy} N.~A., {Pettini} M., 2004, \apj, 607, 226

\bibitem[{{Baldwin}, {Phillips} \& {Terlevich}(1981){Baldwin}, {Phillips}, \&
  {Terlevich}}]{Baldwin:1981}
{Baldwin} J.~A., {Phillips} M.~M., {Terlevich} R., 1981, \pasp, 93, 5

\bibitem[{{Balogh}, {Navarro} \& {Morris}(2000){Balogh}, {Navarro}, \&
  {Morris}}]{Balogh:2000b}
{Balogh} M.~L., {Navarro} J.~F., {Morris} S.~L., 2000, \apj, 540, 113

\bibitem[{{Baugh} {et~al}\mbox{.}(2005){Baugh}, {Lacey}, {Frenk}, {Granato},
  {Silva}, {Bressan}, {Benson}, \& {Cole}}]{Baugh:2005}
{Baugh} C.~M., {Lacey} C.~G., {Frenk} C.~S., {Granato} G.~L., {Silva} L.,
  {Bressan} A., {Benson} A.~J., {Cole} S., 2005, \mnras, 356, 1191

\bibitem[{{Bothwell} {et~al}\mbox{.}(2013){Bothwell}, {Maiolino}, {Kennicutt},
  {Cresci}, {Mannucci}, {Marconi}, \& {Cicone}}]{Bothwell:2013}
{Bothwell} M.~S., {Maiolino} R., {Kennicutt} R., {Cresci} G., {Mannucci} F.,
  {Marconi} A., {Cicone} C., 2013, \mnras, 433, 1425

\bibitem[{{Bouch{\'e}} {et~al}\mbox{.}(2013){Bouch{\'e}}, {Murphy}, {Kacprzak},
  {P{\'e}roux}, {Contini}, {Martin}, \& {Dessauges-Zavadsky}}]{Bouche:2013}
{Bouch{\'e}} N., {Murphy} M.~T., {Kacprzak} G.~G., {P{\'e}roux} C., {Contini}
  T., {Martin} C.~L., {Dessauges-Zavadsky} M., 2013, Science, 341, 50

\bibitem[{{Bournaud}, {Jog} \& {Combes}(2007){Bournaud}, {Jog}, \&
  {Combes}}]{Bournaud:2007}
{Bournaud} F., {Jog} C.~J., {Combes} F., 2007, \aap, 476, 1179

\bibitem[{{Bower}, {Kodama} \& {Terlevich}(1998){Bower}, {Kodama}, \&
  {Terlevich}}]{Bower:1998}
{Bower} R.~G., {Kodama} T., {Terlevich} A., 1998, \mnras, 299, 1193

\bibitem[{{Brinchmann} {et~al}\mbox{.}(2004){Brinchmann}, {Charlot}, {White},
  {Tremonti}, {Kauffmann}, {Heckman}, \& {Brinkmann}}]{Brinchmann:2004}
{Brinchmann} J., {Charlot} S., {White} S.~D.~M., {Tremonti} C., {Kauffmann} G.,
  {Heckman} T., {Brinkmann} J., 2004, \mnras, 351, 1151

\bibitem[{{Brinchmann}, {Kunth} \& {Durret}(2008){Brinchmann}, {Kunth}, \&
  {Durret}}]{Brinchmann:2008a}
{Brinchmann} J., {Kunth} D., {Durret} F., 2008, \aap, 485, 657

\bibitem[{{Brinchmann}, {Pettini} \& {Charlot}(2008){Brinchmann}, {Pettini}, \&
  {Charlot}}]{Brinchmann:2008b}
{Brinchmann} J., {Pettini} M., {Charlot} S., 2008, \mnras, 385, 769

\bibitem[{{Brocklehurst}(1971)}]{Brocklehurst:1971}
{Brocklehurst} M., 1971, \mnras, 153, 471

\bibitem[{{Butcher} \& {Oemler}(1984)}]{Butcher:1984}
{Butcher} H., {Oemler}, Jr. A., 1984, \apj, 285, 426

\bibitem[{{Calzetti} {et~al}\mbox{.}(2000){Calzetti}, {Armus}, {Bohlin},
  {Kinney}, {Koornneef}, \& {Storchi-Bergmann}}]{Calzetti:2000}
{Calzetti} D., {Armus} L., {Bohlin} R.~C., {Kinney} A.~L., {Koornneef} J.,
  {Storchi-Bergmann} T., 2000, \apj, 533, 682

\bibitem[{{Cappellari} {et~al}\mbox{.}(2011){Cappellari}, {Emsellem},
  {Krajnovi{\'c}}, {McDermid}, {Serra}, {Alatalo}, {Blitz}, {Bois}, \&
  et~al.}]{Cappellari:2011}
{Cappellari} M. {et~al.}, 2011, \mnras, 416, 1680

\bibitem[{{Cattaneo} {et~al}\mbox{.}(2008){Cattaneo}, {Dekel}, {Faber}, \&
  {Guiderdoni}}]{Cattaneo:2008}
{Cattaneo} A., {Dekel} A., {Faber} S.~M., {Guiderdoni} B., 2008, \mnras, 389,
  567

\bibitem[{{Chabrier}(2003)}]{Chabrier:2003}
{Chabrier} G., 2003, \pasp, 115, 763

\bibitem[{{Cid Fernandes} {et~al}\mbox{.}(2011){Cid Fernandes},
  {Stasi{\'n}ska}, {Mateus}, \& {Vale Asari}}]{Cid:2011}
{Cid Fernandes} R., {Stasi{\'n}ska} G., {Mateus} A., {Vale Asari} N., 2011,
  \mnras, 413, 1687

\bibitem[{{Coil} {et~al}\mbox{.}(2014){Coil}, {Aird}, {Reddy}, {Shapley},
  {Kriek}, {Siana}, {Mobasher}, {Freeman}, {Price}, \& {Shivaei}}]{Coil:2014}
{Coil} A.~L. {et~al.}, 2014, ArXiv e-prints

\bibitem[{{Cowie} {et~al}\mbox{.}(1996){Cowie}, {Songaila}, {Hu}, \&
  {Cohen}}]{Cowie:1996}
{Cowie} L.~L., {Songaila} A., {Hu} E.~M., {Cohen} J.~G., 1996, \aj, 112, 839

\bibitem[{{Daddi} {et~al}\mbox{.}(2007){Daddi}, {Dickinson}, {Morrison},
  {Chary}, {Cimatti}, {Elbaz}, {Frayer}, {Renzini}, \& et~al.}]{Daddi:2007a}
{Daddi} E. {et~al.}, 2007, \apj, 670, 156

\bibitem[{{Dalcanton}(2007)}]{Dalcanton:2007}
{Dalcanton} J.~J., 2007, \apj, 658, 941

\bibitem[{{Dav{\'e}}, {Finlator} \& {Oppenheimer}(2011){Dav{\'e}}, {Finlator},
  \& {Oppenheimer}}]{Dave:2011b}
{Dav{\'e}} R., {Finlator} K., {Oppenheimer} B.~D., 2011, \mnras, 416, 1354

\bibitem[{{Dekel} {et~al}\mbox{.}(2009){Dekel}, {Birnboim}, {Engel},
  {Freundlich}, {Goerdt}, {Mumcuoglu}, {Neistein}, {Pichon}, {Teyssier}, \&
  {Zinger}}]{Dekel:2009}
{Dekel} A. {et~al.}, 2009, \nat, 457, 451

\bibitem[{{Dopita} {et~al}\mbox{.}(2000){Dopita}, {Kewley}, {Heisler}, \&
  {Sutherland}}]{Dopita:2000}
{Dopita} M.~A., {Kewley} L.~J., {Heisler} C.~A., {Sutherland} R.~S., 2000,
  \apj, 542, 224

\bibitem[{{Dressler} {et~al}\mbox{.}(1997){Dressler}, {Oemler}, {Couch},
  {Smail}, {Ellis}, {Barger}, {Butcher}, {Poggianti}, \&
  {Sharples}}]{Dressler:1997}
{Dressler} A. {et~al.}, 1997, \apj, 490, 577

\bibitem[{{Dressler} {et~al}\mbox{.}(1994){Dressler}, {Oemler}, {Sparks}, \&
  {Lucas}}]{Dressler:1994}
{Dressler} A., {Oemler}, Jr. A., {Sparks} W.~B., {Lucas} R.~A., 1994, \apjl,
  435, L23

\bibitem[{{Dyson} \& {Williams}(1980)}]{Dyson:1980}
{Dyson} J.~E., {Williams} D.~A., 1980, {Physics of the interstellar medium}.
  (Manchester University Press)

\bibitem[{{Ebizuka} {et~al}\mbox{.}(2011){Ebizuka}, {Ichiyama}, {Yamada},
  {Tokoku}, {Onodera}, {Hanesaka}, {Kodate}, {Katsuno Uchimoto}, \&
  et~al.}]{Ebizuka:2011}
{Ebizuka} N. {et~al.}, 2011, \pasj, 63, 605

\bibitem[{{Ellison} {et~al}\mbox{.}(2009){Ellison}, {Simard}, {Cowan},
  {Baldry}, {Patton}, \& {McConnachie}}]{Ellison:2009}
{Ellison} S.~L., {Simard} L., {Cowan} N.~B., {Baldry} I.~K., {Patton} D.~R.,
  {McConnachie} A.~W., 2009, \mnras, 396, 1257

\bibitem[{{Erb}(2008)}]{Erb:2008}
{Erb} D.~K., 2008, \apj, 674, 151

\bibitem[{{Erb} {et~al}\mbox{.}(2010){Erb}, {Pettini}, {Shapley}, {Steidel},
  {Law}, \& {Reddy}}]{Erb:2010}
{Erb} D.~K., {Pettini} M., {Shapley} A.~E., {Steidel} C.~C., {Law} D.~R.,
  {Reddy} N.~A., 2010, \apj, 719, 1168

\bibitem[{{Erb} {et~al}\mbox{.}(2006{\natexlab{a}}){Erb}, {Shapley}, {Pettini},
  {Steidel}, {Reddy}, \& {Adelberger}}]{Erb:2006}
{Erb} D.~K., {Shapley} A.~E., {Pettini} M., {Steidel} C.~C., {Reddy} N.~A.,
  {Adelberger} K.~L., 2006{\natexlab{a}}, \apj, 644, 813

\bibitem[{{Erb} {et~al}\mbox{.}(2006{\natexlab{b}}){Erb}, {Steidel}, {Shapley},
  {Pettini}, {Reddy}, \& {Adelberger}}]{Erb:2006b}
{Erb} D.~K., {Steidel} C.~C., {Shapley} A.~E., {Pettini} M., {Reddy} N.~A.,
  {Adelberger} K.~L., 2006{\natexlab{b}}, \apj, 647, 128

\bibitem[{{Ezer} \& {Cameron}(1971)}]{Ezer:1971}
{Ezer} D., {Cameron} A.~G.~W., 1971, \apss, 14, 399

\bibitem[{{Garn} \& {Best}(2010)}]{Garn:2010}
{Garn} T., {Best} P.~N., 2010, \mnras, 409, 421

\bibitem[{{Gottl{\"o}ber}, {Klypin} \& {Kravtsov}(2001){Gottl{\"o}ber},
  {Klypin}, \& {Kravtsov}}]{Gottlober:2001}
{Gottl{\"o}ber} S., {Klypin} A., {Kravtsov} A.~V., 2001, \apj, 546, 223

\bibitem[{{Hayashi} {et~al}\mbox{.}(2011){Hayashi}, {Kodama}, {Koyama},
  {Tadaki}, \& {Tanaka}}]{Hayashi:2011}
{Hayashi} M., {Kodama} T., {Koyama} Y., {Tadaki} K.-I., {Tanaka} I., 2011,
  \mnras, 415, 2670

\bibitem[{{Hayashi} {et~al}\mbox{.}(2012){Hayashi}, {Kodama}, {Tadaki},
  {Koyama}, \& {Tanaka}}]{Hayashi:2012}
{Hayashi} M., {Kodama} T., {Tadaki} K.-i., {Koyama} Y., {Tanaka} I., 2012,
  \apj, 757, 15

\bibitem[{{Holden} {et~al}\mbox{.}(2014){Holden}, {Oesch}, {Gonzalez},
  {Illingworth}, {Labbe}, {Bouwens}, {Franx}, {van Dokkum}, \&
  et~al.}]{Holden:2014}
{Holden} B.~P. {et~al.}, 2014, ArXiv e-prints

\bibitem[{{Hopkins} {et~al}\mbox{.}(2008){Hopkins}, {Hernquist}, {Cox}, \&
  {Kere{\v s}}}]{Hopkins:2008}
{Hopkins} P.~F., {Hernquist} L., {Cox} T.~J., {Kere{\v s}} D., 2008, \apjs,
  175, 356

\bibitem[{{Hughes} {et~al}\mbox{.}(2013){Hughes}, {Cortese}, {Boselli},
  {Gavazzi}, \& {Davies}}]{Hughes:2013}
{Hughes} T.~M., {Cortese} L., {Boselli} A., {Gavazzi} G., {Davies} J.~I., 2013,
  \aap, 550, A115

\bibitem[{{Ichikawa} {et~al}\mbox{.}(2006){Ichikawa}, {Suzuki}, {Tokoku},
  {Uchimoto}, {Konishi}, {Yoshikawa}, {Yamada}, {Tanaka}, \&
  et~al.}]{Ichikawa:2006}
{Ichikawa} T. {et~al.}, 2006, in Society of Photo-Optical Instrumentation
  Engineers (SPIE) Conference Series, Vol. 6269, Society of Photo-Optical
  Instrumentation Engineers (SPIE) Conference Series

\bibitem[{{Izotov} {et~al}\mbox{.}(2006){Izotov}, {Stasi{\'n}ska}, {Meynet},
  {Guseva}, \& {Thuan}}]{Izotov:2006}
{Izotov} Y.~I., {Stasi{\'n}ska} G., {Meynet} G., {Guseva} N.~G., {Thuan} T.~X.,
  2006, \aap, 448, 955

\bibitem[{{Juneau} {et~al}\mbox{.}(2014){Juneau}, {Bournaud}, {Charlot},
  {Daddi}, {Elbaz}, {Trump}, {Brinchmann}, {Dickinson}, {Duc}, {Gobat},
  {Jean-Baptiste}, {Le Floc'h}, {Lehnert}, {Pacifici}, {Pannella}, \&
  {Schreiber}}]{Juneau:2014}
{Juneau} S. {et~al.}, 2014, \apj, 788, 88

\bibitem[{{Juneau} {et~al}\mbox{.}(2011){Juneau}, {Dickinson}, {Alexander}, \&
  {Salim}}]{Juneau:2011}
{Juneau} S., {Dickinson} M., {Alexander} D.~M., {Salim} S., 2011, \apj, 736,
  104

\bibitem[{{Kashino} {et~al}\mbox{.}(2013){Kashino}, {Silverman}, {Rodighiero},
  {Renzini}, {Arimoto}, {Daddi}, {Lilly}, {Sanders}, \& et~al.}]{Kashino:2013}
{Kashino} D. {et~al.}, 2013, \apjl, 777, L8

\bibitem[{{Kauffmann} {et~al}\mbox{.}(2003){Kauffmann}, {Heckman}, {Tremonti},
  {Brinchmann}, {Charlot}, {White}, {Ridgway}, {Brinkmann}, \&
  et~al.}]{Kauffmann:2003}
{Kauffmann} G. {et~al.}, 2003, \mnras, 346, 1055

\bibitem[{{Kennicutt}(1998)}]{Kennicutt:1998}
{Kennicutt}, Jr. R.~C., 1998, \araa, 36, 189

\bibitem[{{Kewley} \& {Dopita}(2002)}]{Kewley:2002}
{Kewley} L.~J., {Dopita} M.~A., 2002, \apjs, 142, 35

\bibitem[{{Kewley} {et~al}\mbox{.}(2013{\natexlab{a}}){Kewley}, {Dopita},
  {Leitherer}, {Dav{\'e}}, {Yuan}, {Allen}, {Groves}, \&
  {Sutherland}}]{Kewley:2013b}
{Kewley} L.~J., {Dopita} M.~A., {Leitherer} C., {Dav{\'e}} R., {Yuan} T.,
  {Allen} M., {Groves} B., {Sutherland} R., 2013{\natexlab{a}}, \apj, 774, 100

\bibitem[{{Kewley} {et~al}\mbox{.}(2001){Kewley}, {Dopita}, {Sutherland},
  {Heisler}, \& {Trevena}}]{Kewley:2001}
{Kewley} L.~J., {Dopita} M.~A., {Sutherland} R.~S., {Heisler} C.~A., {Trevena}
  J., 2001, \apj, 556, 121

\bibitem[{{Kewley} \& {Ellison}(2008)}]{Kewley:2008}
{Kewley} L.~J., {Ellison} S.~L., 2008, \apj, 681, 1183

\bibitem[{{Kewley} {et~al}\mbox{.}(2006){Kewley}, {Groves}, {Kauffmann}, \&
  {Heckman}}]{Kewley:2006}
{Kewley} L.~J., {Groves} B., {Kauffmann} G., {Heckman} T., 2006, \mnras, 372,
  961

\bibitem[{{Kewley} {et~al}\mbox{.}(2013{\natexlab{b}}){Kewley}, {Maier},
  {Yabe}, {Ohta}, {Akiyama}, {Dopita}, \& {Yuan}}]{Kewley:2013a}
{Kewley} L.~J., {Maier} C., {Yabe} K., {Ohta} K., {Akiyama} M., {Dopita} M.~A.,
  {Yuan} T., 2013{\natexlab{b}}, \apjl, 774, L10

\bibitem[{{Kobulnicky} \& {Kewley}(2004)}]{Kobulnicky:2004}
{Kobulnicky} H.~A., {Kewley} L.~J., 2004, \apj, 617, 240

\bibitem[{{Kodama} {et~al}\mbox{.}(2004){Kodama}, {Balogh}, {Smail}, {Bower},
  \& {Nakata}}]{Kodama:2004}
{Kodama} T., {Balogh} M.~L., {Smail} I., {Bower} R.~G., {Nakata} F., 2004,
  \mnras, 354, 1103

\bibitem[{{Kodama} \& {Bower}(2001)}]{Kodama:2001}
{Kodama} T., {Bower} R.~G., 2001, \mnras, 321, 18

\bibitem[{{Kodama} {et~al}\mbox{.}(2013){Kodama}, {Hayashi}, {Koyama},
  {Tadaki}, {Tanaka}, \& {Shimakawa}}]{Kodama:2013}
{Kodama} T., {Hayashi} M., {Koyama} Y., {Tadaki} K.-i., {Tanaka} I.,
  {Shimakawa} R., 2013, in IAU Symposium, Vol. 295, IAU Symposium, {Thomas} D.,
  {Pasquali} A., {Ferreras} I., eds., pp. 74--77

\bibitem[{{Kodama} {et~al}\mbox{.}(2007){Kodama}, {Tanaka}, {Kajisawa}, {Kurk},
  {Venemans}, {De Breuck}, {Vernet}, \& {Lidman}}]{Kodama:2007}
{Kodama} T., {Tanaka} I., {Kajisawa} M., {Kurk} J., {Venemans} B., {De Breuck}
  C., {Vernet} J., {Lidman} C., 2007, \mnras, 377, 1717

\bibitem[{{Koyama} {et~al}\mbox{.}(2010){Koyama}, {Kodama}, {Shimasaku},
  {Hayashi}, {Okamura}, {Tanaka}, \& {Tokoku}}]{Koyama:2010}
{Koyama} Y., {Kodama} T., {Shimasaku} K., {Hayashi} M., {Okamura} S., {Tanaka}
  I., {Tokoku} C., 2010, \mnras, 403, 1611

\bibitem[{{Koyama} {et~al}\mbox{.}(2014){Koyama}, {Kodama}, {Tadaki},
  {Hayashi}, {Tanaka}, \& {Shimakawa}}]{Koyama:2014}
{Koyama} Y., {Kodama} T., {Tadaki} K.-i., {Hayashi} M., {Tanaka} I.,
  {Shimakawa} R., 2014, ArXiv e-prints

\bibitem[{{Koyama} {et~al}\mbox{.}(2013{\natexlab{a}}){Koyama}, {Kodama},
  {Tadaki}, {Hayashi}, {Tanaka}, {Smail}, {Tanaka}, \& {Kurk}}]{Koyama:2013a}
{Koyama} Y., {Kodama} T., {Tadaki} K.-i., {Hayashi} M., {Tanaka} M., {Smail}
  I., {Tanaka} I., {Kurk} J., 2013{\natexlab{a}}, \mnras, 428, 1551

\bibitem[{{Koyama} {et~al}\mbox{.}(2013{\natexlab{b}}){Koyama}, {Smail},
  {Kurk}, {Geach}, {Sobral}, {Kodama}, {Nakata}, {Swinbank}, {Best}, {Hayashi},
  \& {Tadaki}}]{Koyama:2013b}
{Koyama} Y. {et~al.}, 2013{\natexlab{b}}, \mnras, 434, 423

\bibitem[{{Krabbe} {et~al}\mbox{.}(2014){Krabbe}, {Rosa}, {Dors}, {Pastoriza},
  {Winge}, {H{\"a}gele}, {Cardaci}, \& {Rodrigues}}]{Krabbe:2014}
{Krabbe} A.~C., {Rosa} D.~A., {Dors} O.~L., {Pastoriza} M.~G., {Winge} C.,
  {H{\"a}gele} G.~F., {Cardaci} M.~V., {Rodrigues} I., 2014, \mnras, 437, 1155

\bibitem[{{Kriss}(1994)}]{Kriss:1994}
{Kriss} G., 1994, Astronomical Data Analysis Software and Systems, 3, 437

\bibitem[{{Kulas} {et~al}\mbox{.}(2013){Kulas}, {McLean}, {Shapley}, {Steidel},
  {Konidaris}, {Matthews}, {Mace}, {Rudie}, \& et~al.}]{Kulas:2013}
{Kulas} K.~R. {et~al.}, 2013, \apj, 774, 130

\bibitem[{{Kurk} {et~al}\mbox{.}(2004){Kurk}, {Pentericci}, {R{\"o}ttgering},
  \& {Miley}}]{Kurk:2004a}
{Kurk} J.~D., {Pentericci} L., {R{\"o}ttgering} H.~J.~A., {Miley} G.~K., 2004,
  \aap, 428, 793

\bibitem[{{Kurk} {et~al}\mbox{.}(2000){Kurk}, {R{\"o}ttgering}, {Pentericci},
  {Miley}, {van Breugel}, {Carilli}, {Ford}, {Heckman}, \& et~al.}]{Kurk:2000}
{Kurk} J.~D. {et~al.}, 2000, \aap, 358, L1

\bibitem[{{Larson}(1981)}]{Larson:1981}
{Larson} R.~B., 1981, \mnras, 194, 809

\bibitem[{{Levesque} \& {Richardson}(2014)}]{Levesque:2014}
{Levesque} E.~M., {Richardson} M.~L.~A., 2014, \apj, 780, 100

\bibitem[{{Maeder}(1987)}]{Maeder:1987}
{Maeder} A., 1987, \aap, 178, 159

\bibitem[{{Magrini} {et~al}\mbox{.}(2012){Magrini}, {Sommariva}, {Cresci},
  {Sani}, {Galametz}, {Mannucci}, {Petropoulou}, \& {Fumana}}]{Magrini:2012}
{Magrini} L., {Sommariva} V., {Cresci} G., {Sani} E., {Galametz} A., {Mannucci}
  F., {Petropoulou} V., {Fumana} M., 2012, \mnras, 426, 1195

\bibitem[{{Maiolino} {et~al}\mbox{.}(2008){Maiolino}, {Nagao}, {Grazian},
  {Cocchia}, {Marconi}, {Mannucci}, {Cimatti}, {Pipino}, \&
  et~al.}]{Maiolino:2008}
{Maiolino} R. {et~al.}, 2008, \aap, 488, 463

\bibitem[{{Mannucci} {et~al}\mbox{.}(2010){Mannucci}, {Cresci}, {Maiolino},
  {Marconi}, \& {Gnerucci}}]{Mannucci:2010}
{Mannucci} F., {Cresci} G., {Maiolino} R., {Marconi} A., {Gnerucci} A., 2010,
  \mnras, 408, 2115

\bibitem[{{Mannucci} {et~al}\mbox{.}(2009){Mannucci}, {Cresci}, {Maiolino},
  {Marconi}, {Pastorini}, {Pozzetti}, {Gnerucci}, {Risaliti}, {Schneider},
  {Lehnert}, \& {Salvati}}]{Mannucci:2009}
{Mannucci} F. {et~al.}, 2009, \mnras, 398, 1915

\bibitem[{{Mannucci}, {Salvaterra} \& {Campisi}(2011){Mannucci}, {Salvaterra},
  \& {Campisi}}]{Mannucci:2011}
{Mannucci} F., {Salvaterra} R., {Campisi} M.~A., 2011, \mnras, 414, 1263

\bibitem[{{Masters} {et~al}\mbox{.}(2014){Masters}, {McCarthy}, {Siana},
  {Malkan}, {Mobasher}, {Atek}, {Henry}, {Martin}, {Rafelski}, {Hathi},
  {Scarlata}, {Ross}, {Bunker}, {Blanc}, {Bedregal}, {Dom{\'{\i}}nguez},
  {Colbert}, {Teplitz}, \& {Dressler}}]{Masters:2014}
{Masters} D. {et~al.}, 2014, \apj, 785, 153

\bibitem[{{McCarthy} {et~al}\mbox{.}(2008){McCarthy}, {Frenk}, {Font}, {Lacey},
  {Bower}, {Mitchell}, {Balogh}, \& {Theuns}}]{McCarthy:2008}
{McCarthy} I.~G., {Frenk} C.~S., {Font} A.~S., {Lacey} C.~G., {Bower} R.~G.,
  {Mitchell} N.~L., {Balogh} M.~L., {Theuns} T., 2008, \mnras, 383, 593

\bibitem[{{McGaugh}(1991)}]{Mcgaugh:1991}
{McGaugh} S.~S., 1991, \apj, 380, 140

\bibitem[{{McLean} {et~al}\mbox{.}(2012){McLean}, {Steidel}, {Epps},
  {Konidaris}, {Matthews}, {Adkins}, {Aliado}, {Brims}, {Canfield}, {Cromer},
  {Fucik}, {Kulas}, {Mace}, {Magnone}, {Rodriguez}, {Rudie}, {Trainor}, {Wang},
  {Weber}, \& {Weiss}}]{McLean:2012}
{McLean} I.~S. {et~al.}, 2012, in Society of Photo-Optical Instrumentation
  Engineers (SPIE) Conference Series, Vol. 8446, Society of Photo-Optical
  Instrumentation Engineers (SPIE) Conference Series

\bibitem[{{Mihos} \& {Hernquist}(1996)}]{Mihos:1996}
{Mihos} J.~C., {Hernquist} L., 1996, \apj, 464, 641

\bibitem[{{Nagao}, {Maiolino} \& {Marconi}(2006){Nagao}, {Maiolino}, \&
  {Marconi}}]{Nagao:2006}
{Nagao} T., {Maiolino} R., {Marconi} A., 2006, \aap, 459, 85

\bibitem[{{Nagashima} {et~al}\mbox{.}(2005{\natexlab{a}}){Nagashima}, {Lacey},
  {Baugh}, {Frenk}, \& {Cole}}]{Nagashima:2005a}
{Nagashima} M., {Lacey} C.~G., {Baugh} C.~M., {Frenk} C.~S., {Cole} S.,
  2005{\natexlab{a}}, \mnras, 358, 1247

\bibitem[{{Nagashima} {et~al}\mbox{.}(2005{\natexlab{b}}){Nagashima}, {Lacey},
  {Okamoto}, {Baugh}, {Frenk}, \& {Cole}}]{Nagashima:2005b}
{Nagashima} M., {Lacey} C.~G., {Okamoto} T., {Baugh} C.~M., {Frenk} C.~S.,
  {Cole} S., 2005{\natexlab{b}}, \mnras, 363, L31

\bibitem[{{Nakajima} \& {Ouchi}(2014)}]{Nakajima:2014}
{Nakajima} K., {Ouchi} M., 2014, \mnras, 442, 900

\bibitem[{{Nakajima} {et~al}\mbox{.}(2013){Nakajima}, {Ouchi}, {Shimasaku},
  {Hashimoto}, {Ono}, \& {Lee}}]{Nakajima:2013}
{Nakajima} K., {Ouchi} M., {Shimasaku} K., {Hashimoto} T., {Ono} Y., {Lee}
  J.~C., 2013, \apj, 769, 3

\bibitem[{{Newman} {et~al}\mbox{.}(2014){Newman}, {Buschkamp}, {Genzel},
  {F{\"o}rster Schreiber}, {Kurk}, {Sternberg}, {Gnat}, {Rosario}, \&
  et~al.}]{Newman:2014}
{Newman} S.~F. {et~al.}, 2014, \apj, 781, 21

\bibitem[{{Newman} {et~al}\mbox{.}(2012){Newman}, {Shapiro Griffin}, {Genzel},
  {Davies}, {F{\"o}rster-Schreiber}, {Tacconi}, {Kurk}, {Wuyts}, \&
  et~al.}]{Newman:2012}
{Newman} S.~F. {et~al.}, 2012, \apj, 752, 111

\bibitem[{{Okamoto} \& {Nagashima}(2003)}]{Okamoto:2003}
{Okamoto} T., {Nagashima} M., 2003, \apj, 587, 500

\bibitem[{{Oppenheimer} \& {Dav{\'e}}(2008)}]{Oppenheimer:2008}
{Oppenheimer} B.~D., {Dav{\'e}} R., 2008, \mnras, 387, 577

\bibitem[{{Peng} \& {Maiolino}(2014)}]{Peng:2014}
{Peng} Y.-j., {Maiolino} R., 2014, \mnras, 438, 262

\bibitem[{{Pentericci} {et~al}\mbox{.}(2002){Pentericci}, {Kurk}, {Carilli},
  {Harris}, {Miley}, \& {R{\"o}ttgering}}]{Pentericci:2002}
{Pentericci} L., {Kurk} J.~D., {Carilli} C.~L., {Harris} D.~E., {Miley} G.~K.,
  {R{\"o}ttgering} H.~J.~A., 2002, \aap, 396, 109

\bibitem[{{Pettini} \& {Pagel}(2004)}]{Pettini:2004}
{Pettini} M., {Pagel} B.~E.~J., 2004, \mnras, 348, L59

\bibitem[{{Prieto} {et~al}\mbox{.}(2013){Prieto}, {Eliche-Moral}, {Balcells},
  {Crist{\'o}bal-Hornillos}, {Erwin}, {Abreu}, {Dom{\'{\i}}nguez-Palmero},
  {Hempel}, \& et~al.}]{Prieto:2013}
{Prieto} M. {et~al.}, 2013, \mnras, 428, 999

\bibitem[{{Puech} {et~al}\mbox{.}(2006){Puech}, {Flores}, {Hammer}, \&
  {Lehnert}}]{Puech:2006}
{Puech} M., {Flores} H., {Hammer} F., {Lehnert} M.~D., 2006, \aap, 455, 131

\bibitem[{{Salim} {et~al}\mbox{.}(2007){Salim}, {Rich}, {Charlot},
  {Brinchmann}, {Johnson}, {Schiminovich}, {Seibert}, {Mallery}, {Heckman},
  {Forster}, {Friedman}, {Martin}, {Morrissey}, {Neff}, {Small}, {Wyder},
  {Bianchi}, {Donas}, {Lee}, {Madore}, {Milliard}, {Szalay}, {Welsh}, \&
  {Yi}}]{Salim:2007}
{Salim} S. {et~al.}, 2007, \apjs, 173, 267

\bibitem[{{Salpeter}(1955)}]{Salpeter:1955}
{Salpeter} E.~E., 1955, \apj, 121, 161

\bibitem[{{Sanders} {et~al}\mbox{.}(2014){Sanders}, {Shapley}, {Kriek},
  {Reddy}, {Freeman}, {Coil}, {Siana}, {Mobasher}, {Shivaei}, {Price}, \& {de
  Groot}}]{Sanders:2014}
{Sanders} R.~L. {et~al.}, 2014, ArXiv e-prints

\bibitem[{{Shapley} {et~al}\mbox{.}(2005{\natexlab{a}}){Shapley}, {Coil}, {Ma},
  \& {Bundy}}]{Shapley:2005}
{Shapley} A.~E., {Coil} A.~L., {Ma} C.-P., {Bundy} K., 2005{\natexlab{a}},
  \apj, 635, 1006

\bibitem[{{Shapley} {et~al}\mbox{.}(2014){Shapley}, {Reddy}, {Kriek},
  {Freeman}, {Sanders}, {Siana}, {Coil}, {Mobasher}, {Shivaei}, {Price}, \& {de
  Groot}}]{Shapley:2014}
{Shapley} A.~E. {et~al.}, 2014, ArXiv e-prints

\bibitem[{{Shapley} {et~al}\mbox{.}(2005{\natexlab{b}}){Shapley}, {Steidel},
  {Erb}, {Reddy}, {Adelberger}, {Pettini}, {Barmby}, \&
  {Huang}}]{Shapley:2005b}
{Shapley} A.~E., {Steidel} C.~C., {Erb} D.~K., {Reddy} N.~A., {Adelberger}
  K.~L., {Pettini} M., {Barmby} P., {Huang} J., 2005{\natexlab{b}}, \apj, 626,
  698

\bibitem[{{Shimakawa} {et~al}\mbox{.}(2014{\natexlab{a}}){Shimakawa}, {Kodama},
  {Steidel}, {Tadaki}, {Tanaka}, {Strom}, {Hayashi}, {Koyama}, {Suzuki}, \&
  {Yamamoto}}]{Shimakawa:2014c}
{Shimakawa} R. {et~al.}, 2014{\natexlab{a}}, ArXiv e-prints

\bibitem[{{Shimakawa} {et~al}\mbox{.}(2014{\natexlab{b}}){Shimakawa}, {Kodama},
  {Tadaki}, {Tanaka}, {Hayashi}, \& {Koyama}}]{Shimakawa:2014}
{Shimakawa} R., {Kodama} T., {Tadaki} K.-i., {Tanaka} I., {Hayashi} M.,
  {Koyama} Y., 2014{\natexlab{b}}, \mnras, 441, L1

\bibitem[{{Shirazi}, {Brinchmann} \& {Rahmati}(2014){Shirazi}, {Brinchmann}, \&
  {Rahmati}}]{Shirazi:2014}
{Shirazi} M., {Brinchmann} J., {Rahmati} A., 2014, \apj, 787, 120

\bibitem[{{Smail} {et~al}\mbox{.}(2014){Smail}, {Geach}, {Swinbank}, {Tadaki},
  {Arumugam}, {Hartley}, {Almaini}, {Bremer}, \& et~al.}]{Smail:2014}
{Smail} I. {et~al.}, 2014, \apj, 782, 19

\bibitem[{{Springel} {et~al}\mbox{.}(2005){Springel}, {White}, {Jenkins},
  {Frenk}, {Yoshida}, {Gao}, {Navarro}, {Thacker}, {Croton}, {Helly},
  {Peacock}, {Cole}, {Thomas}, {Couchman}, {Evrard}, {Colberg}, \&
  {Pearce}}]{Springel:2005}
{Springel} V. {et~al.}, 2005, \nat, 435, 629

\bibitem[{{Steidel} {et~al}\mbox{.}(2010){Steidel}, {Erb}, {Shapley},
  {Pettini}, {Reddy}, {Bogosavljevi{\'c}}, {Rudie}, \& {Rakic}}]{Steidel:2010}
{Steidel} C.~C., {Erb} D.~K., {Shapley} A.~E., {Pettini} M., {Reddy} N.,
  {Bogosavljevi{\'c}} M., {Rudie} G.~C., {Rakic} O., 2010, \apj, 717, 289

\bibitem[{{Steidel} {et~al}\mbox{.}(2014){Steidel}, {Rudie}, {Strom},
  {Pettini}, {Reddy}, {Shapley}, {Trainor}, {Erb}, {Turner}, {Konidaris},
  {Kulas}, {Mace}, {Matthews}, \& {McLean}}]{Steidel:2014}
{Steidel} C.~C. {et~al.}, 2014, ArXiv e-prints

\bibitem[{{Steidel} {et~al}\mbox{.}(2004){Steidel}, {Shapley}, {Pettini},
  {Adelberger}, {Erb}, {Reddy}, \& {Hunt}}]{Steidel:2004}
{Steidel} C.~C., {Shapley} A.~E., {Pettini} M., {Adelberger} K.~L., {Erb}
  D.~K., {Reddy} N.~A., {Hunt} M.~P., 2004, \apj, 604, 534

\bibitem[{{Stott} {et~al}\mbox{.}(2013){Stott}, {Sobral}, {Bower}, {Smail},
  {Best}, {Matsuda}, {Hayashi}, {Geach}, \& {Kodama}}]{Stott:2013}
{Stott} J.~P. {et~al.}, 2013, \mnras, 436, 1130

\bibitem[{{Suzuki} {et~al}\mbox{.}(2008){Suzuki}, {Tokoku}, {Ichikawa},
  {Uchimoto}, {Konishi}, {Yoshikawa}, {Tanaka}, {Yamada}, \&
  et~al.}]{Suzuki:2008}
{Suzuki} R. {et~al.}, 2008, \pasj, 60, 1347

\bibitem[{{Tadaki} {et~al}\mbox{.}(2014){Tadaki}, {Kodama}, {Tamura},
  {Hayashi}, {Koyama}, {Shimakawa}, {Tanaka}, {Kohno}, {Hatsukade}, \&
  {Suzuki}}]{Tadaki:2014b}
{Tadaki} K.-i. {et~al.}, 2014, ArXiv e-prints

\bibitem[{{Tadaki} {et~al}\mbox{.}(2013){Tadaki}, {Kodama}, {Tanaka},
  {Hayashi}, {Koyama}, \& {Shimakawa}}]{Tadaki:2013}
{Tadaki} K.-i., {Kodama} T., {Tanaka} I., {Hayashi} M., {Koyama} Y.,
  {Shimakawa} R., 2013, \apj, 778, 114

\bibitem[{{Thomas} {et~al}\mbox{.}(2005){Thomas}, {Maraston}, {Bender}, \&
  {Mendes de Oliveira}}]{Thomas:2005}
{Thomas} D., {Maraston} C., {Bender} R., {Mendes de Oliveira} C., 2005, \apj,
  621, 673

\bibitem[{{Tremonti} {et~al}\mbox{.}(2004){Tremonti}, {Heckman}, {Kauffmann},
  {Brinchmann}, {Charlot}, {White}, {Seibert}, {Peng}, \&
  et~al.}]{Tremonti:2004}
{Tremonti} C.~A. {et~al.}, 2004, \apj, 613, 898

\bibitem[{{Troncoso} {et~al}\mbox{.}(2014){Troncoso}, {Maiolino}, {Sommariva},
  {Cresci}, {Mannucci}, {Marconi}, {Meneghetti}, {Grazian}, \&
  et~al.}]{Troncoso:2014}
{Troncoso} P. {et~al.}, 2014, \aap, 563, A58

\bibitem[{{Valentino} {et~al}\mbox{.}(2014){Valentino}, {Daddi}, {Strazzullo},
  {Gobat}, {Onodera}, {Bournaud}, {Juneau}, {Renzini}, {Arimoto}, {Carollo}, \&
  {Zanella}}]{Valentino:2014}
{Valentino} F. {et~al.}, 2014, ArXiv e-prints

\bibitem[{{Veilleux} \& {Osterbrock}(1987)}]{Veilleux:1987}
{Veilleux} S., {Osterbrock} D.~E., 1987, \apjs, 63, 295

\bibitem[{{Whitaker} {et~al}\mbox{.}(2014){Whitaker}, {Rigby}, {Brammer},
  {Gladders}, {Sharon}, {Teng}, \& {Wuyts}}]{Whitaker:2014}
{Whitaker} K.~E., {Rigby} J.~R., {Brammer} G.~B., {Gladders} M.~D., {Sharon}
  K., {Teng} S.~H., {Wuyts} E., 2014, ArXiv e-prints

\bibitem[{{Whitaker} {et~al}\mbox{.}(2012){Whitaker}, {van Dokkum}, {Brammer},
  \& {Franx}}]{Whitaker:2012}
{Whitaker} K.~E., {van Dokkum} P.~G., {Brammer} G., {Franx} M., 2012, \apjl,
  754, L29

\bibitem[{{Wuyts} {et~al}\mbox{.}(2014){Wuyts}, {Rigby}, {Gladders}, \&
  {Sharon}}]{Wuyts:2014}
{Wuyts} E., {Rigby} J.~R., {Gladders} M.~D., {Sharon} K., 2014, \apj, 781, 61

\bibitem[{{Wuyts} {et~al}\mbox{.}(2013){Wuyts}, {F{\"o}rster Schreiber},
  {Nelson}, {van Dokkum}, {Brammer}, {Chang}, {Faber}, {Ferguson}, {Franx}, \&
  et~al.}]{Wuyts:2013}
{Wuyts} S. {et~al.}, 2013, \apj, 779, 135

\bibitem[{{Yabe} {et~al}\mbox{.}(2014{\natexlab{a}}){Yabe}, {Ohta}, {Akiyama},
  {Iwamuro}, {Tamura}, {Yuma}, {Dalton}, \& {Lewis}}]{Yabe:2014b}
{Yabe} K., {Ohta} K., {Akiyama} M., {Iwamuro} F., {Tamura} N., {Yuma} S.,
  {Dalton} G., {Lewis} I., 2014{\natexlab{a}}, ArXiv e-prints

\bibitem[{{Yabe} {et~al}\mbox{.}(2014{\natexlab{b}}){Yabe}, {Ohta}, {Iwamuro},
  {Akiyama}, {Tamura}, {Yuma}, {Kimura}, {Takato}, {Moritani}, {Sumiyoshi},
  {Maihara}, {Silverman}, {Dalton}, {Lewis}, {Bonfield}, {Lee}, {Curtis-Lake},
  {Macaulay}, \& {Clarke}}]{Yabe:2014}
{Yabe} K. {et~al.}, 2014{\natexlab{b}}, \mnras, 437, 3647

\bibitem[{{Yabe} {et~al}\mbox{.}(2012){Yabe}, {Ohta}, {Iwamuro}, {Yuma},
  {Akiyama}, {Tamura}, {Kimura}, {Takato}, \& et~al.}]{Yabe:2012}
{Yabe} K. {et~al.}, 2012, \pasj, 64, 60

\bibitem[{{Yoshikawa} {et~al}\mbox{.}(2010){Yoshikawa}, {Akiyama}, {Kajisawa},
  {Alexander}, {Ohta}, {Suzuki}, {Tokoku}, {Uchimoto}, \&
  et~al.}]{Yoshikawa:2010}
{Yoshikawa} T. {et~al.}, 2010, \apj, 718, 112

\bibitem[{{Zahid} {et~al}\mbox{.}(2014{\natexlab{a}}){Zahid}, {Dima},
  {Kudritzki}, {Kewley}, {Geller}, {Hwang}, {Silverman}, \&
  {Kashino}}]{Zahid:2014}
{Zahid} H.~J., {Dima} G.~I., {Kudritzki} R.-P., {Kewley} L.~J., {Geller} M.~J.,
  {Hwang} H.~S., {Silverman} J.~D., {Kashino} D., 2014{\natexlab{a}}, \apj,
  791, 130

\bibitem[{{Zahid} {et~al}\mbox{.}(2013){Zahid}, {Geller}, {Kewley}, {Hwang},
  {Fabricant}, \& {Kurtz}}]{Zahid:2013}
{Zahid} H.~J., {Geller} M.~J., {Kewley} L.~J., {Hwang} H.~S., {Fabricant}
  D.~G., {Kurtz} M.~J., 2013, \apjl, 771, L19

\bibitem[{{Zahid} {et~al}\mbox{.}(2014{\natexlab{b}}){Zahid}, {Kashino},
  {Silverman}, {Kewley}, {Daddi}, {Renzini}, {Rodighiero}, {Nagao}, {Arimoto},
  {Sanders}, {Kartaltepe}, {Lilly}, {Maier}, {Geller}, {Capak}, {Carollo},
  {Chu}, {Hasinger}, {Ilbert}, {Kajisawa}, {Koekemoer}, {Kovac{\#728}}, {Le
  F{\`e}vre}, {Masters}, {McCracken}, {Onodera}, {Scoville}, {Strazzullo},
  {Sugiyama}, {Taniguchi}, \& {The COSMOS Team}}]{Zahid:2014b}
{Zahid} H.~J. {et~al.}, 2014{\natexlab{b}}, \apj, 792, 75

\end{thebibliography}

\appendix
\section{Dust extinction across the mass--metallicity 
relation in SDSS galaxies}

In order to discuss the effect of possible sampling bias 
on the derived gaseous metallicities between the UV-selected 
galaxies (which are strongly affected by dust extinction) 
and the \ha\ selected galaxies, we here investigate the 
correlation between the amount of dust extinction and the 
deviation from the M--Z relation in the SDSS SF galaxies.

\begin{figure}
\centering
\includegraphics[width=75mm]{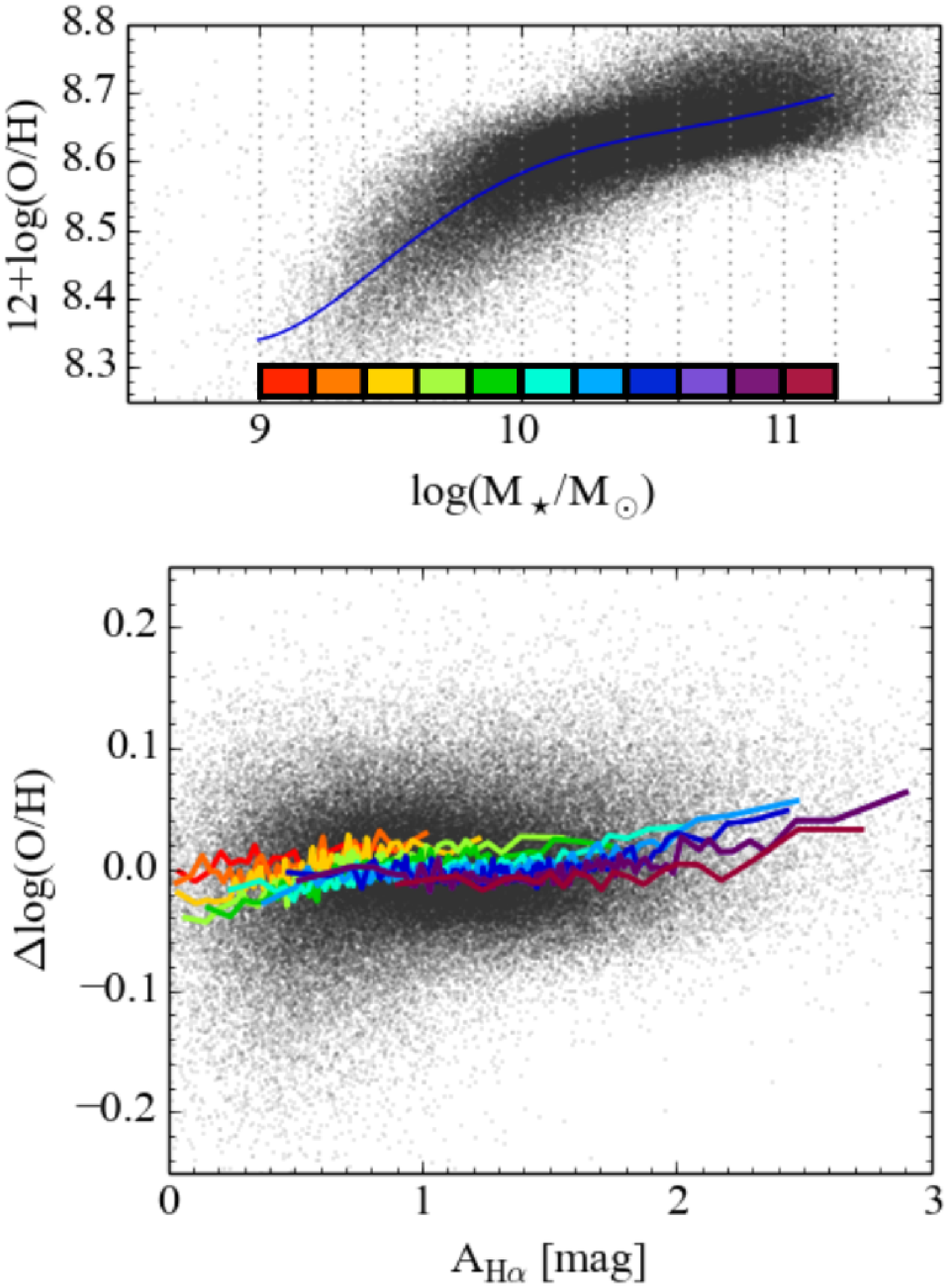}
\caption{(a) The upper panel shows the mass--metallicity relation
of SDSS galaxies based on the N2 index \citep{Pettini:2004}.
(b) The lower panel represents the offset of the metallicity 
($\Delta$ log (O/H)) from the median locus of the M--Z relation
shown in the upper panel (blue curve) as a function of
A$_\mathrm{H\alpha}$ derived from the Balmer decrement. 
Colour-coded lines indicate median values of 
$\Delta$log(O/H) as a function of A$_\mathrm{H\alpha}$
for the subsamples divided according to their stellar masses
as shown in the vertical dotted lines in the upper panel and the
corresponding colour coding.}
\label{fig10}
\end{figure}

For this purpose, we select H$\alpha$ emitters in the
SDSS in the same way as we use in this paper (see \S2.1).
Fig. \ref{fig10}a shows the M--Z relation of these galaxies
obtained using the N2 index \citep{Pettini:2004}.
Fig. \ref{fig10}b represents the deviations in metallicity from
the M--Z relation as a function of dust obscuration in H$\alpha$
line.
We divide the sample into 11 bins according to their stellar
masses (Fig. \ref{fig10}a), and plot the locus of median values
of $\Delta$ log (O/H) as a function of A$_\mathrm{H\alpha}$
for each mass bin in a different colour (Fig. \ref{fig10}b). 
As a result, we do not find any strong correlation between these
two quantities, meaning that the effect of dust obscuration
in the UV selected sample does not introduce any significant
bias to the derived metallicities.
Therefore we conclude that the selection effect we discuss in
\S4.2 should be minimal.

%%%%%%%%%%%%%
%%%%%%%%%%%%%

\label{lastpage}

\end{document}